\documentclass[a4paper,fleqn]{cas-dc}

\usepackage{subcaption}
\usepackage[numbers]{natbib}
\usepackage{datetime}
\newdateformat{mydate}{\THEDAY\ \monthname[\THEMONTH]\ \THEYEAR}
\mydate
\usepackage{tikz}
\newcommand{\rev}[1]{#1}
\usetikzlibrary{arrows.meta, positioning}
\usepackage{enumitem}
\usepackage{placeins}

\def\tsc#1{\csdef{#1}{\textsc{\lowercase{#1}}\xspace}}
\tsc{WGM}
\tsc{QE}
\newdefinition{definition}{Definition}
\newproof{pf}{Proof}

\begin{document}
\let\WriteBookmarks\relax
\def\floatpagepagefraction{1}
\def\textpagefraction{.001}

\shorttitle{Nonlinear DPC Methods for Residential Heating}

\shortauthors{Zieglmeier et al.}

\title[mode = title]{Model-Free Control for Residential Heating: Deployment and Simulation of Nonlinear Data-Enabled Predictive Control}
\tnotemark[1]

% Title footnote
\tnotetext[1]{S.~Zieglmeier was supported by UiO:Energy and Environment at the University of Oslo and acknowledges Empa, the Swiss Federal Laboratories for Materials Science and Technology, for providing the research environment used in this work. C.~Verhoek acknowledges the support from the European COVER project (grant No.~101086228). M.~Hudoba de Badyn is supported by FME Solar, funded by the Research Council of Norway under project number 350244.
%\\
%Code available: \url{https://github.com/SebsDevLab/Nonlinear_DPC_heating}
}

%% -------------------------------------------------------
%% AUTHORS
%% -------------------------------------------------------

% First author: Sebastian Zieglmeier
\author[1,2]{Sebastian Zieglmeier}[
  orcid=0009-0009-2516-3723]
\cormark[1]
\ead{sebastiz@uio.no, seb.zieglmeier@gmail.com}
\credit{Conceptualization, Data curation, Formal analysis, Funding acquisition, Investigation, Methodology, Software, Validation, Visualization, Writing – original draft}

\affiliation[1]{organization={Department of Technology Systems, University of Oslo},
            %addressline={},
            city={Kjeller},
            postcode={2027},
            %state={},
            country={Norway}}
            
\affiliation[2]{organization={Urban Energy Systems Laboratory, Swiss Federal Laboratories for Materials Science and Technology (Empa)},
            %addressline={},
            city={Duebendorf},
            postcode={8600},
            %state={},
            country={Switzerland}}
% Second author: Chris Verhoek
\author[3, 4]{Chris Verhoek}[orcid=0000-0001-6633-6103]
\ead{cverhoek@seas.upenn.edu, c.verhoek@tue.nl}
\credit{Supervision, Writing -- Review \& Editing}

\affiliation[3]{organization={Department of Electrical and Systems Engineering, University of Pennsylvania},
            %addressline={},
            city={Philadelphia},
            postcode={19104},
            state={PA},
            country={United States}}

% Note: C. Verhoek also affiliated with TU/e
\affiliation[4]{organization={Control Systems Group, Eindhoven University of Technology},
            %addressline={},
            city={Eindhoven},
            postcode={5600 MB},
            %state={},
            country={The Netherlands}}

% Third author: Jaap Eising
\author[5]{Jaap Eising}[orcid=0000-0003-2155-8196]
\ead{j.eising@rug.nl}
\credit{Supervision, Writing -- Review \& Editing}

\affiliation[5]{organization={Engineering and Technology Institute, University of Groningen},
            %addressline={},
            city={Groningen},
            postcode={9712},
            %state={},
            country={The Netherlands}}

% Fourth author: Mathias Hudoba de Badyn
\author[1]{Mathias {Hudoba de Badyn}}[orcid=0000-0003-0955-2381]
\ead{mathihud@uio.no}
\credit{Supervision, Funding acquisition, Writing -- Review \& Editing}

\cortext[1]{Corresponding author}
%% -------------------------------------------------------
%% ABSTRACT
%% -------------------------------------------------------
\begin{abstract}
Residential heating accounts for a large share of building energy use, and predictive control strategies can reduce it by anticipating rather than reacting to the room temperature alone. The usual manner of implementing predictive control, model predictive control (MPC), requires an accurate model of each individual residential unit for heating control. Obtaining this model is time-consuming to obtain manually or even impossible, and such models do not transfer across a heterogeneous building stock. Data-enabled predictive control (DeePC) removes this modeling step by designing the controller directly using measured trajectories of the system. The foundations of DeePC, however, are built on the class of deterministic linear time-invariant systems, which is an unrealistic assumption for residential heating. 
This work applies three recently proposed nonlinear extensions of DeePC, collectively referred to as data-driven predictive control (DPC): Select-DPC, gain-scheduling DPC, and linear parameter-varying DPC. 
These are compared against standard DeePC and the hysteresis controller, the industry standard in residential heating. The comparison runs over a full heating season on a calibrated digital twin of an occupied research unit, the NEST research building in Switzerland, assessed by heating energy and comfort-band violation.
% Moreover, gain-scheduling DPC was deployed on the real residential apartment and validated the conclusions from the simulations. Each of the DPC controllers consumes significantly less energy than the hysteresis controller, amounting to roughly $11\%$ over the season, validating the use of DPC. 
% Changed to: 
\rev{Moreover, gain-scheduling DPC was deployed on the real residential apartment, where it kept the room temperature within the comfort band and showed the same qualitative behavior as in simulation. Each of the DPC controllers consumes less energy than the hysteresis controller over the season, at the cost of higher comfort-band violations. With a tightened lower comfort bound, Select-DPC saves $10.7\%$ of the heating energy while also violating the comfort band less than the hysteresis controller.}
The nonlinear methods further outperform linear DeePC in all assessed scenarios, and we analyze the relative costs and benefits of each nonlinear method.
\end{abstract}


% Keywords

\maketitle

%% -------------------------------------------------------
%% SECTION 1: INTRODUCTION
%% -------------------------------------------------------
\section{Introduction}\label{sec:intro}
The building sector is among the largest single contributors to global energy use and greenhouse-gas emissions, responsible for roughly 30\% of global final energy consumption and a comparable share of energy-related carbon emissions~\citep{iea2023buildings}. Of this, space and water heating constitute the dominant end use, and in the residential sector heating alone accounts for the large majority of household energy demand~\citep{iea2023buildings}. This makes residential heating a central target for decarbonization. With the building stock turning over slowly, and deep retrofits remaining costly, the efficient operation of existing heating systems has emerged as one of the most immediately scalable solutions for reducing energy consumption and emissions~\citep{stoffel2023evaluation}.

Much of this potential is lost because conventional control is purely reactive. Hysteresis-type control is the standard in residential heating for good reason: it is robust, requires neither a model nor tuning, and deploys straightforwardly across the wide variety of system configurations found across the building stock.
Yet a residential unit is continuously exposed to time-varying external conditions, such as the ambient outdoor temperature, which largely dictates how much heating is actually required. Passive heat gains, for instance from occupants, add to this and are uncertain and difficult to predict. Solar radiation adds a further, nonlinear contribution, varying with the position of the sun relative to the building~\citep{bunning2022physics}.
A controller that switches reactively on the current temperature alone is blind to these influences, and may, for example, keep injecting heat that the apartment was already about to receive from the environment. Realizing the available savings, therefore, requires control that considers these disturbances in the decision-making rather than merely reacting to them~\citep{drgovna2020all}.

Model predictive control (MPC) is the established advanced alternative, computing the control action at each step by solving a constrained optimization problem that uses disturbance forecasts and explicit constraint handling to deliver such savings~\citep{oldewurtel2012use}.
Its central weakness is the model, since accurate first-principles or gray-box models are building-specific, degrade as the building changes over time, and are labor-intensive or even impossible to obtain~\citep{drgovna2020all, 7087366, vzavcekova2014towards}.
This weakness is compounded by the sheer heterogeneity of the residential sector, where units differ in construction, configuration, occupancy pattern, and local environment, among countless other factors. No single model generalizes across this variation, so each building would require its manually built and maintained model, which prevents MPC from scaling in residential heating~\citep{7087366}.

Data-driven predictive control (DPC) sidesteps this bottleneck, replacing the modeling step with the direct use of data. Data-enabled predictive control (DeePC)~\citep{coulson2019data} is one such approach, which builds upon a theoretical foundation that allows stability and robustness guarantees to be provided~\citep{berberich2020data, berberich2021data}.
The theoretical basis is Willems' fundamental lemma~\citep{willems2005note}, which states that for a linear time-invariant (LTI) system, a single sufficiently rich measured trajectory spans all trajectories of a certain length that the system can produce. 
The recorded data itself can therefore directly serve as the predictor within the predictive controller, in place of a difficult-to-obtain parametric model.
Nevertheless, Willems' fundamental lemma restricts the approach to deterministic LTI systems. Regularization terms within the optimization problem of the controller can provide some measure of robustness against noise and nonlinearities, but address these only to a limited extent~\citep{giacomelli2025insights, zieglmeier2025semi}.
Therefore, thermal building dynamics, which are inherently nonlinear, e.g., through operating point-dependent solar heat gains, call for more advanced data-driven predictive control (DPC) methods that adapt the data-driven predictor to the current operating conditions. 
In this study, we consider three such methods, namely gain-scheduling DPC (GS-DPC), Select-DPC, and linear parameter-varying DPC (LPV-DPC), which localize the predictor by scheduling between region-specific data sets, by selecting the most relevant past trajectories online, and through a measurable parameter-varying embedding, respectively. These methods are compared with standard DeePC and the hysteresis controller that is industry standard in residential heating~\citep{svetozarevic2022data}. The comparison is carried out on an occupied apartment of the Next Evolution in Sustainable Building Technologies (NEST) research building in Switzerland~\citep{richner2017nest}, first on its calibrated digital twin and then on the physical unit itself.

\subsection{Related Work}
Classic MPC with a parametric model was applied extensively to buildings, with experimental studies on building climate control reported in~\citep{oldewurtel2012use, vsiroky2011experimental, 7087366, vzavcekova2014towards}. On NEST, MPC has been benchmarked against rule-based control on the digital twin employed here~\citep{khayatian2022benchmarking} and deployed on the residential unit for emission-aware operation~\citep{cai2024experimental}. Gaussian-process MPC constitutes a further data-driven MPC method, deployed on the HVAC system of a hospital~\citep{maddalena2022experimental}. As research on MPC for building climate control has grown considerably over the past two decades, we refer to~\citep{drgovna2020all}, which gives a comprehensive overview of its formulations, modeling paradigms, and implementation aspects. In all of these, the parametric model remains the bottleneck, since it is building-specific and does not transfer across the heterogeneous stock.

Data-driven building control has been pursued along several lines. One retains the MPC framework and replaces its parametric model by one learned from data, e.g. by using random forests~\citep{BUNNING2020109792}, neural networks~\citep{bunning2021input, yang2020model}, or physics-informed linear regression~\citep{bunning2022physics}. Reinforcement learning learns a control policy from the data instead of solving a (non-convex) optimization problem online, and has been developed for room temperature control on units of the NEST demonstrator in~\citep{di2021deep, di2022near, svetozarevic2022data}. Both require substantial amounts of data or an accurate simulator to train against, and the resulting models and policies are often difficult to interpret.

The data can also be used directly in the controller, without an intermediate model or policy. DPCs built on Willems' fundamental lemma follow this route, and the optimization remains convex. It has recently been applied to real buildings.
A bilevel reformulation of DeePC, in which the trajectory prediction enters as a lower-level problem within a robust control problem, and the data are updated online, was validated experimentally on a single-zone office building~\citep{10089206}.
This bilevel DeePC was later compared against Gaussian-process MPC and deep reinforcement learning, each deployed on a different building. They concluded that it is the most straightforward of the three to set up, but applicable only to systems that are at most approximately linear~\citep{9992445}.
Signal matrix model predictive control, an indirect data-driven extension of DeePC, was benchmarked against DeePC, its bilevel variant, and a hysteresis controller for space heating on the same unit and the same digital twin used in this work, and subsequently deployed on the physical unit~\citep{yin2024data}.
In each of the aforementioned works, the nonlinearity of the building is either left to regularization or absorbed into an additional output-error term, rather than addressed by the predictor itself.

A range of nonlinear extensions of DPC have recently been proposed in the control literature. In this work, we focus on Select-DPC~\citep{beerwerth2025morecontextualsamplingnonlinear, naf2025choose}, GS-DPC~\citep{guerrero2025gain, zieglmeier2025gain} and LPV-DPC~\citep{verhoek2021data, verhoek2025linear}. Like DeePC, all three build the predictor from data collected offline. The approach of~\citet{berberich2022linear} instead updates it online, which requires the conditions on the data to hold continuously and presumes slowly evolving dynamics.
Two alternative methods in the same vein appeared after the simulation study presented here was carried out:~\citep{engeln2026data} localizes the predictor by weighting the data columns according to their distance to the current operating point, retaining the full data matrix rather than selecting a subset, and~\citep{giacomelli2026beyond} extends the fundamental lemma to piecewise affine systems.
Lifting-based nonlinear DeePC variants, such as Koopman or kernelized formulations~\citep{huang2023robust, lazar2024basis, lian2021nonlinear, lian2021koopman}, were not considered, as they require the design of a suitable lifting a priori, a choice for which no systematic procedure exists yet and which reintroduces a plant-specific modeling effort that the considered direct data-driven methods avoid.

\subsection{Contributions}
The main contributions of this paper are as follows:
\begin{enumerate}[label={C\arabic*:}, align=left, ref={C\arabic*}, leftmargin=*]
\item We adapt the considered DPC frameworks to residential heating. The control problem is formulated such that the measurable disturbances, the ambient outdoor temperature, and the solar radiation are explicitly accounted for in the data-driven predictor, while the remaining influences enter implicitly through the data sequences. \label{contrib:dpcs}
\item We conduct a direct, systematic comparison of the nonlinear methods against standard DeePC and the hysteresis controller on a high-fidelity digital twin of the Next Evolution in Sustainable Building Technologies (NEST) research building in Switzerland.
The digital twin is driven throughout by real measurements recorded at the building, and the controllers are operated over a full heating season, spanning varying weather conditions. The performance is evaluated in terms of energy consumption and thermal comfort.
\item We deploy one of the approaches, GS-DPC, on the NEST building, demonstrating that the method functions reliably under real-world operating conditions and that the behavior observed in simulation carries over to the physical unit.
\end{enumerate}
The remainder of the paper is organized as follows. Section~\ref{sec:application} describes the high-fidelity simulation environment and the real-world experimental setting. Section~\ref{sec:methods} introduces the controllers and their modifications to the original formulation. Sections~\ref{sec:sim-results} and~\ref{sec:exp-results} present the simulation and experimental results, respectively, and Section~\ref{sec:conclusion} concludes this work.

\subsection{Notation}
The set of real numbers is denoted by $\mathbb{R}$, and $\mathbb{N}_{[a,\,b]} := \{a,\,a+1,\,\dots,\,b\}$ is the set of consecutive integers from $a$ to $b$. For a symmetric positive (semi-)definite matrix $Q$ we write $Q \succ 0$ ($Q \succeq 0$), and $\|x\|_Q^2 := x^\top Q x$ denotes the corresponding weighted squared norm, and $\|\cdot\|_2$ denotes the Euclidean norm. The stacking operator $\operatorname{col}(v_1,\dots,v_n) := [\,v_1^\top\ \cdots\ v_n^\top\,]^\top$ vertically concatenates its arguments, and $\otimes$ denotes the Kronecker product. For a discrete-time signal $w$, the stacked vector of its samples over $\mathbb{N}_{[a,\,b]}$ is written $w_{[a,\,b]} := \operatorname{col}(w_a,\dots,w_b)$. Historic data is indicated with the superscript $d$, e.g., $w^d$. Given a data sequence $w^d = (w^d_0,\dots,w^d_{T-1})$ with $w^d_k \in \mathbb{R}^{n_w}$, its depth-$L$ Hankel matrix is
\begin{equation}
H_L(w^d) := \begin{bmatrix} w^d_0 & w^d_1 & \cdots & w^d_{T-L} \\ w^d_1 & w^d_2 & \cdots & w^d_{T-L+1} \\ \vdots & \vdots & \ddots & \vdots \\ w^d_{L-1} & w^d_L & \cdots & w^d_{T-1} \end{bmatrix} \in \mathbb{R}^{n_w L \times (T-L+1)}. \label{eq:hankel_def}
\end{equation}
A sequence $w^d$ is \emph{persistently exciting} of order $L$ if its Hankel matrix $H_L(w^d)$ has full row rank.

\section{Application: Residential Heating}\label{sec:application}
\begin{figure}
    \centering
    \includegraphics[width=0.75\linewidth]{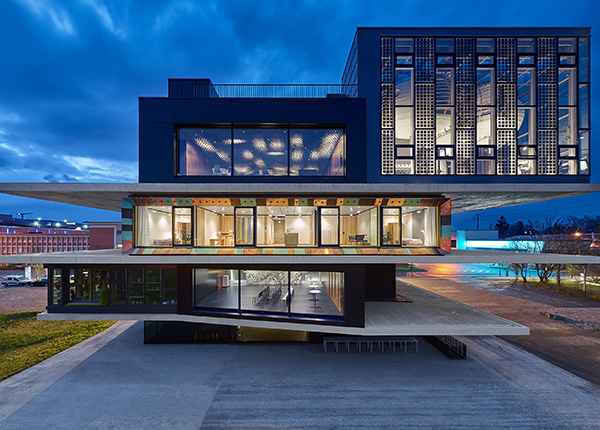}
    \caption{The NEST building with the UMAR unit in the center of the image. \copyright Zooey Braun.}
    \label{fig:NEST}
\end{figure}

\begin{figure}
    \centering
    \includegraphics[width=0.75\linewidth]{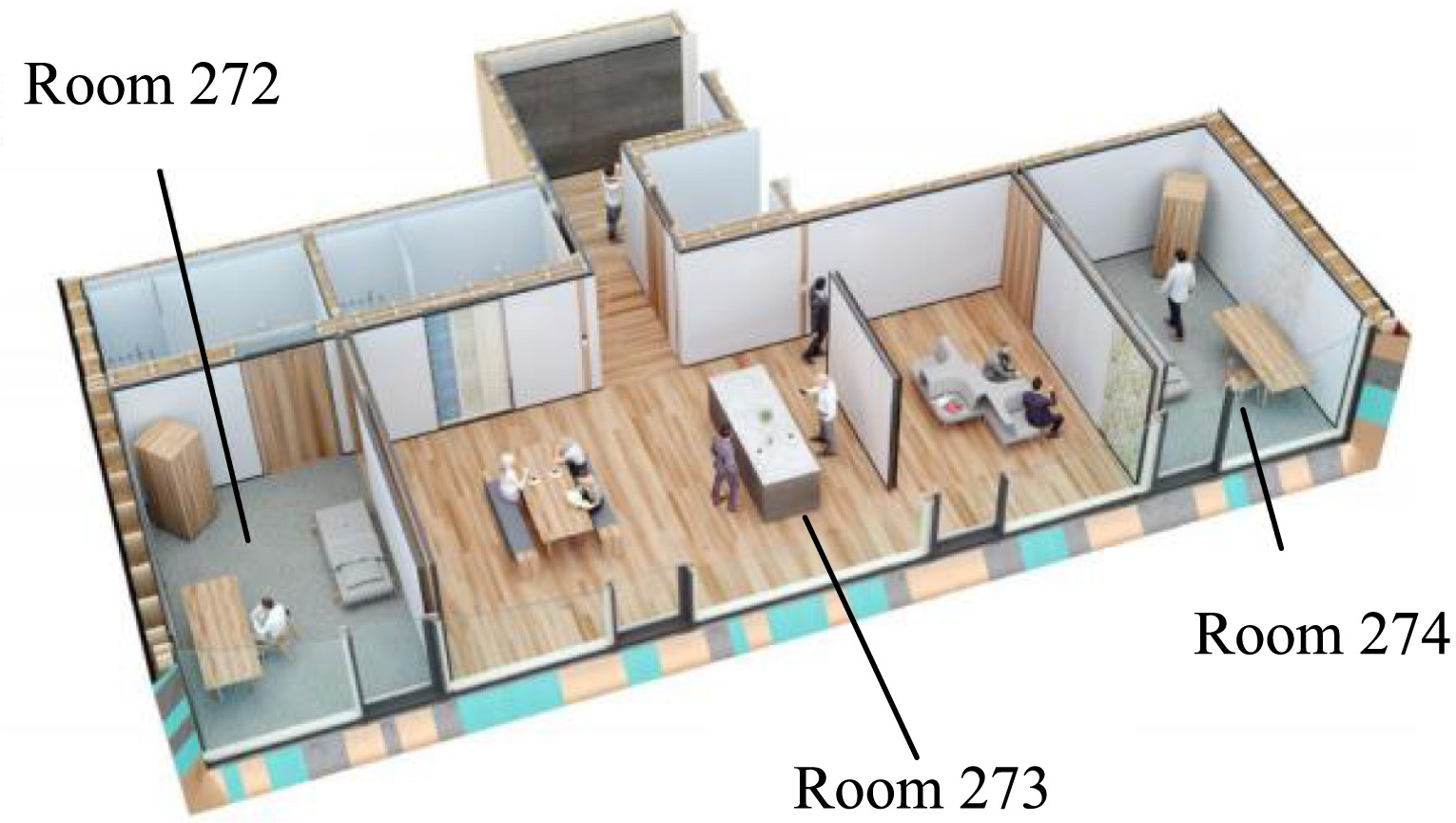}
    \caption{Layout of the UMAR unit with the three controlled rooms. \copyright Werner Sobek.}
    \label{fig:UMAR}
\end{figure}

The study is based on the Urban Mining and Recycling (UMAR) unit of the NEST building (Figure~\ref{fig:NEST}) at Empa in Dübendorf, Switzerland, an occupied residential apartment used as a research demonstrator~\citep{heisel2019resource, richner2017nest}, shown in Figures~\ref{fig:NEST} and~\ref{fig:UMAR}. A calibrated digital twin of the unit exists, built on an EnergyPlus model and driven by measurements recorded at the building, which reproduces its thermal behavior at high fidelity~\citep{bojarski2023nestli, khayatian2022benchmarking}. The real unit defines the experimental setup, while the digital twin permits the repeatable, direct comparison platform that the physical building cannot provide.


\subsection{The UMAR Unit}\label{sec:umar}
UMAR comprises a living room, two bedrooms, and two bathrooms. The two bedrooms, Room 272 (R272) and Room 274 (R274), and the living room, Room 273 (R273), marked in Figure~\ref{fig:UMAR}, are heated and controlled in this work. Heating power serves as the input and is delivered through ceiling radiant panels supplied from a hot-water grid operating at roughly $35\,^\circ\mathrm{C}$, while the room temperature is the controlled output.

The thermal behavior of these rooms depends on the interaction between the supplied heat, the building's thermal storage and losses, and a large number of external influences. These include the outdoor air temperature, the solar radiation entering through the windows, internal gains from occupants, electrical appliances, air exchange, infiltration, wind, stack effect, window operation, long-wave radiative exchange, ground-coupled conduction through the floor, thermal coupling to adjacent rooms, the time-varying supply temperature of the hot-water grid, and humidity.
Each of these acts on its own timescale and with its own magnitude, but the system's thermal mass yields slow room temperature dynamics, that is, the room responds to both heating and external influences with substantial inertia and lag. None of these factors is individually negligible, but several are neither separately measurable nor directly attributable to a single physical cause, which makes their individual effect on the room temperature difficult to identify and model.
Nevertheless, two factors dominate the heat balance and are both measurable and forecast-able, which is what allows them to be anticipated by a predictive controller. First, the outdoor air temperature sets the heat lost through the building envelope, together with the heat carried in by infiltrating outdoor air, and thereby the baseline heating power required to hold a given room temperature. Secondly, solar radiation enters through the windows as a substantial and strongly time-varying passive heat gain. Therefore, the effects of outdoor temperature and solar radiation are treated explicitly within this work.

Three properties of this system are central to the choice of control approach:
\begin{enumerate}
\item \textbf{Complexity:} The heat balance is shaped by a multitude of simultaneous, heterogeneous disturbances, most of which cannot be measured in isolation, so that a controller must deal robustly with the effects it cannot individually identify or model.

\item \textbf{Poor scalability:} The relative weighting of these influences differs from one unit to the next according to its construction, orientation, occupancy, and surroundings, so that any model identified for one unit does not transfer directly to another. Per-unit parametric modeling therefore scales poorly across the heterogeneous building stock.

\item \textbf{Nonlinearity:} The heat a given level of solar radiation injects into a room depends on the sun's position relative to the window, since the angle of incidence sets the transmitted fraction. Solar radiation is measured by a weather station on the roof, which records the global irradiance available across the open sky rather than the fraction penetrating a given window. The same measured irradiance therefore heats a room differently depending on the time of day and the season, introducing nonlinear behavior.
\end{enumerate}
Taken together, these three properties motivate a shared solution: the direct use of data for controller design.

\subsection{High-Fidelity Digital Twin}\label{sec:digital-twin}
A direct comparison of several controllers on the physical unit is not feasible: each controller would necessarily run over a different period, under different weather, occupancy, and operating conditions. Hence, the observed performance differences could not be attributed to the control law rather than to the environment. We therefore conduct the comparative study on \textit{nestli}~\citep{bojarski2023nestli}, a digital twin of the UMAR unit, which reproduces the same environmental conditions for every controller and thereby allows us to isolate the effect of the control law~\citep{khayatian2022benchmarking}.

The digital twin is built on an EnergyPlus model of the UMAR unit~\citep{crawley2001energyplus}, which resolves the influences identified above individually. These influences are based on real measurements from the building, so it is exercised under the weather, occupancy, and operating conditions actually observed rather than under synthetic ones. The model is calibrated against multi-year on-site measurements recorded at a one-minute resolution and runs at the same resolution. It is wrapped as a Functional Mock-up Unit (FMU) and orchestrated within a co-simulation framework, and has been validated as a benchmarking environment for building automation and control~\citep{khayatian2022benchmarking}.
This level of detail is what justifies treating it as a high-fidelity stand-in for the real unit, since it captures the dominant and secondary influences, and the dynamics that characterize the physical system, rather than an overly simplified first-principles model, such as a resistance-capacitance model~\citep{sturzenegger2014brcm}.

With the system and its high-fidelity twin established, the environment is in place to address the central question of this paper. Does addressing the nonlinearity of the system and incorporating the weather forecast in the prediction pay off, and how do the nonlinear DPC approaches compare against the hysteresis controller and against each other?

\section{Methods}\label{sec:methods}
An efficient approach to manage nonlinearities in dynamical systems is to employ the class of linear parameter-varying (LPV) systems~\citep{verhoek2024encyclo}. This class considers systems with a linear input-(state)-output relationship, while this relationship itself varies along a measurable \emph{scheduling signal}~$\rho$. The scheduling signal captures all the nonlinearities of the system, and thus dictates the current (local, linear) operating condition. To be precise, a discrete-time nonlinear system whose nonlinearities are captured using a time-varying signal $\rho_k$, such that we can represent the system by the following LPV state-space realization:
\begin{comment}
% "Original" version (Less LPV/control theory focus)
Consider a discrete-time nonlinear system whose nonlinearities are captured using a time-varying signal $\rho_k$, %called the scheduling variable, such that the resulting system representation is \emph{linear parameter-varying}:
\end{comment}
%
\begin{subequations}\label{eq:nl_statespace}
\begin{align}
x_{k+1} &= A(\rho_k)\,x_k + B(\rho_k)\,u_k + E(\rho_k)\,d_k + \eta_k, \\
y_k &= C(\rho_k)\,x_k + D(\rho_k)\,u_k + v_k,
\end{align}
\end{subequations}
with state $x_k \in \mathbb{R}^{n_x}$, control input $u_k \in \mathbb{R}^{n_u}$, measured disturbance $d_k \in \mathbb{R}^{n_d}$, output $y_k \in \mathbb{R}^{n_y}$, unknown residual term $\eta_k \in \mathbb{R}^{n_x}$, and output noise $v_k \in \mathbb{R}^{n_y}$. The scheduling-dependent system matrices are $A(\rho_k) \in \mathbb{R}^{n_x \times n_x}$, $B(\rho_k) \in \mathbb{R}^{n_x \times n_u}$, $E(\rho_k) \in \mathbb{R}^{n_x \times n_d}$, $C(\rho_k) \in \mathbb{R}^{n_y \times n_x}$, and $D(\rho_k) \in \mathbb{R}^{n_y \times n_u}$. Note that for a fixed $\rho_k^\star$, the system is LTI and represents the local linear behavior of the original nonlinear system around the operating point defined by $\rho_k^\star$.

For the residential heating system of Section~\ref{sec:application}, the signals of~\eqref{eq:nl_statespace} have a direct physical meaning. 
% The control input $u_k$ is the heating power $P_{\mathrm{h},k}$ delivered to a room, and the output $y_k$ is its measured air temperature $T_{\mathrm{room},k}$. 
% Changed to: 
\rev{The control input $u_k$ is the heating power computed by the controller, which the inner loop of Section~\ref{sec:control-setup} realizes as the delivered heating power $P_{\mathrm{h},k}$. The recorded data and the initial condition of the predictive controllers contain the delivered power $P_{\mathrm{h},k}$. The output $y_k$ is the measured air temperature $T_{\mathrm{room},k}$ of the room.}
The disturbance collects the dominant external influences identified in Section~\ref{sec:application}, the outdoor air temperature and the solar irradiance, $d_k = \operatorname{col}(T_{\mathrm{out},k}, I_{\mathrm{sol},k})$. Both are measured at the building and available as a forecast. The state $x_k$ is the heat stored in the building's thermal masses, which evolves slowly and is not measured.

The matrices $A(\rho_k),\dots,E(\rho_k)$ vary with the scheduling variable $\rho_k$ and thereby capture the nonlinearity, so that each value of $\rho_k$ corresponds to a different operating point, while the residual $\eta_k$ collects the unmeasured residual effects explained in Section~\ref{sec:application}. The operating point dependence of the disturbance gain $E(\rho_k)$ reflects that a given measured radiation enters a room differently depending on the sun's position.
Therefore, we take the scheduling variable $\rho_k$ to be the solar azimuth relative to the building, an exogenous signal that traces the sun's position and is exactly known at any time from the location and time of day.

Representation~\eqref{eq:nl_statespace} is not identified from data or used directly by any of the control methods below. It instead illustrates, in familiar control terms, the nonlinearity and complexity these methods have to address. The control methods in the following make use of this structure to different degrees. Some exploit the scheduling dependence explicitly, while others retain a single global representation, treating the system as approximately linear and leaning on regularization to absorb what that approximation leaves out.

\subsection{Control Architecture and Objective}\label{sec:control-setup}
\begin{figure*}
\centering
\begin{tikzpicture}[
    scale=1, transform shape,
    >={Stealth[length=2.4mm]},
    font=\small,
    block/.style={draw, rounded corners, minimum width=2.3cm, minimum height=1.15cm, align=center},
    line/.style={draw, ->},
    dot/.style={circle, fill, inner sep=1.3pt}
]
% --- loop boundaries (drawn first, so they sit behind) ---
\draw[blue!70, dashed, thick, rounded corners] (3.2,-2) rectangle (10.2, 1.4);
\node[blue!70, anchor=south west, font=\footnotesize] at (3.2,0.9) {inner loop, $16.7$\,mHz};
\draw[red!70, dashed, thick, rounded corners] (-1.5,-2.95) rectangle (11.2,1.6);
\node[red!70, anchor=south west, font=\footnotesize] at (-1.4,-3.0) {outer loop, $1.1$\,mHz};
% --- main forward chain ---
\node[block] (outer) at (0,0) {Outer-loop\\controller};
\node[block] (inner) at (4.8,0) {Inner loop:\\power PWM};
\node[draw, rounded corners, fill=white, inner sep=3pt] (plant) at (8.3,0)
      {\includegraphics[width=3.1cm]{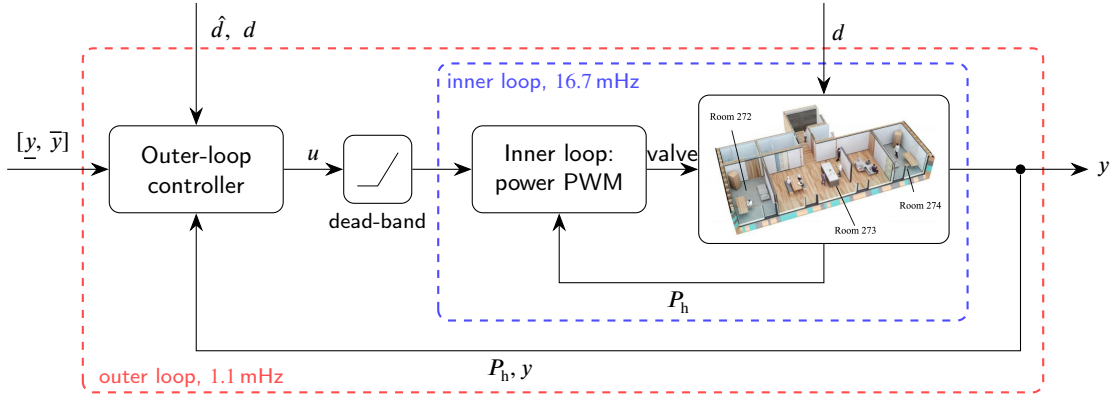}};
% --- dead-band block between the loops ---
\node[draw, rounded corners, minimum width=0.9cm, minimum height=0.9cm, inner sep=1pt, fill=white] (db) at (2.4,0) {};
\draw[very thin] (2.13,-0.22) -- (2.4,-0.22) -- (2.67,0.22);
\node[anchor=north, font=\footnotesize] at (db.south) {dead-band};
% --- reference into the outer controller ---
\draw[line] (-2.5,0) -- node[above, xshift=-2mm]{$[\underline{y},\,\overline{y}]$} (outer.west);
% --- commanded heating power through the dead-band ---
\draw[line] (outer.east) -- node[above]{$u$} (db.west);
\draw[line] (db.east) -- (inner.west);
% --- valve actuation ---
\draw[line] (inner.east) -- node[above]{valve} (plant.west);
% --- measured output ---
\node[dot] (p1) at (10.9,0) {};
\draw (plant.east) -- (11.6,0);
\draw[line] (11.6,0) -- (11.8,0) node[right]{$y$};
% --- disturbance into the plant ---
\draw[line] (8.3,2.2) -- node[above, xshift=2mm]{$d$} (plant.north);
% --- forecast and past disturbances into the outer controller ---
\draw[line] (0,2.2) -- node[above, xshift=5mm, yshift=2mm]{$\hat{d},\ d$} (outer.north);
% --- inner loop feedback ---
\draw[line] (plant.south) -- (8.3,-1.5) -- (4.8,-1.5)
            node[pos=0.55,below]{$P_\mathrm{h}$} -- (inner.south);
% --- outer loop feedback ---
\draw[line] (p1) -- (10.9,-2.4) -- (0,-2.4)
            node[pos=0.62,below]{$P_\mathrm{h}, y$} -- (outer.south);
\end{tikzpicture}
\caption{Two-layer control architecture. The outer-loop controller, instantiated by the hysteresis controller or one of the DPCs, computes a commanded heating power $u$ from the comfort band, the room temperature $y$, and optionally disturbance forecasts and measurements. The actuation dead-band sets $u$ to zero below $\varepsilon P_{\max}$, and the inner loop realizes the remaining command by pulse-width modulation of the valve.}
\label{fig:control-architecture}
\end{figure*}

All controllers in this work share a two-layer architecture that separates the predictive decision from its physical realization, as illustrated in Figure~\ref{fig:control-architecture}. An outer loop runs at a fixed control interval of $15$\,min, corresponding to $1.1$\,mHz, and decides, for each room, how much heat to deliver over the next interval, expressed as a heating power $u_k$. The inner loop then realizes this commanded power on the radiant panel at $16.7$\,mHz. Because the panel is fed from the hot-water grid at a varying supply temperature, a fixed valve opening does not correspond to a fixed thermal power. The inner loop therefore modulates the valve by pulse-width modulation, adjusting the fraction of the interval for which it is open so that the time-averaged delivered power $P_{\mathrm{h},k}$ matches the commanded $u_k$.
Since NEST operates on a one-minute resolution, valve openings below one minute cannot be realized, so even an arbitrarily small commanded power would open the valve for a full minute. An actuation dead-band is therefore applied between the two loops, setting $u_k$ to zero whenever it falls below a fraction $\varepsilon$ of the maximum power $P_{\max}$. The same resolution limits how precisely the delivered power can follow the commanded one, so $P_{\mathrm{h},k}$ deviates from the computed $u_k$.

The outer loop is the object of the control design in this work. It is filled by one of the controllers introduced below, the hysteresis controller or one of the four data-driven ones, which differ only in how $u_k$ is computed. The two-layer architecture lets every controller reason in terms of a single, physically meaningful quantity, the delivered heating power, while the inner loop absorbs the actuator and supply-side dynamics. The inner loop is part of the building automation design at NEST rather than a design choice of this work.
Each of the three controlled rooms is regulated by its own independent controller, with its own data matrices, optimization problem, and hyperparameter set. The thermal coupling between rooms is not covered explicitly, and enters for every room as a disturbance $\eta$.

All predictive controllers share the same objective: minimize the heating energy while keeping each room thermally comfortable.
The comfort temperature of the room is specified not as a fixed setpoint but as a band $[\underline{y}_k,\,\overline{y}_k]$, within which any temperature is acceptable. Tracking a band rather than a reference leaves the controller free to choose where within the band to operate, which is the freedom it exploits to save energy. The bounds are time-varying, following an occupancy schedule that admits a wider band when the unit is unoccupied~\citep{yin2024data}.
The band is penalized in the cost rather than enforced as a hard constraint. A room can leave the band through effects the controller cannot counteract, such as solar overheating or an open window, and a hard constraint would render the optimization infeasible whenever this occurs. Comfort violations therefore enter as a penalty on the amount by which the temperature leaves the band, through the lower and upper violations
\begin{equation}
\underline{s}_k = \max(0,\, \underline{y}_k - y_k), \qquad \overline{s}_k = \max(0,\, y_k - \overline{y}_k). \label{eq:band_slack}
\end{equation}
The decision variable is the commanded heating power $u_k$, which is bounded by the available panel power, $0 \le u_k \le P_{\max}$, and enters as a constraint. Over a prediction horizon of length $N$, the cost function used in the DPCs is
\begin{equation}
\ell(u, y) = \sum_{k=0}^{N-1} \Big( \|\underline{s}_k\|_Q^2 + \|\overline{s}_k\|_Q^2 + \|u_k\|_R^2 \Big), \label{eq:cost}
\end{equation}
with the design weights $Q \succ 0$ on the comfort violations and $R \succ 0$ on the heating power. The ratio of $Q$ to $R$ sets the trade-off between thermal comfort and energy use. 
% Since the delivered power over an interval of fixed length $\Delta k$ corresponds to the supplied heating energy $E=P_{\mathrm{h}}\,\Delta k$, penalizing the power minimizes the consumed energy, which is the quantity reported in the results.
% Changed to: 
\rev{The heating energy supplied over an interval of length $\Delta k$ is $E=P_{\mathrm{h},k}\,\Delta k$. Since $u_k \ge 0$, penalizing the heating power reduces the supplied energy. The quadratic penalty, the standard choice in receding-horizon control, renders the optimization strictly convex and yields a smooth heating power. It is not identical to the supplied energy, which is linear in the power, so the predictive controllers do not minimize the energy directly. All controllers are nevertheless evaluated on the same quantity, the delivered heating energy defined in Section~\ref{subsec:results}.}

\subsection{Hysteresis Controller}\label{sec:hysteresis}
As a benchmark, we use a hysteresis controller, which is standard in residential heating. It is purely reactive, switching the heat on when the room temperature falls to a lower bound ($y_k<\underline{y}_k$) and off again once it reaches an upper bound ($\overline{y}_k<y_k$), with no model, no forecast, and no tuning beyond the bounds themselves. It uses neither the disturbance measurements nor any prediction, and therefore cannot anticipate the heat the building is about to receive for free or to lose to the environment. It provides a reactive, industry-standard reference for energy consumption against which the predictive controllers are measured.

\subsection{Data-Enabled Predictive Control}\label{sec:deepc}
Classical MPC would require identifying a parametric model of the form~\eqref{eq:nl_statespace}. DeePC instead replaces that model with a single, sufficiently rich sequence of measured data~\citep{coulson2019data}, building the predictor directly from a trajectory of the unit with no explicit model parameters, zone capacities, or thermal resistances to identify. The heterogeneous influences on the room temperature, including those never measured, are not modeled individually but enter implicitly through the recorded data, in the combinations and magnitudes in which they naturally occur in the building. Transferring the controller to another unit requires only a new (sufficiently rich) data sequence rather than the development of a whole new model. This makes the approach scalable across a building stock for which parametric modeling does not transfer. Throughout this work, DeePC denotes the original formulation from~\citep{coulson2019data} for LTI systems, while DPC refers to the overarching family of data-driven control methods inspired by it, including the nonlinear methods we consider in this work and we discuss in the following subsections.

Let $(u^d, y^d)$ denote an input--output trajectory of length $T$ collected offline, from which we form the Hankel matrices $H_L(u^d)$ and $H_L(y^d)$ of depth $L = T_\mathrm{ini} + N$. We assume the input $u^d$ is persistently exciting, which together with controllability is sufficient, though not necessary, for these Hankel matrices to span the behavior of a linear time-invariant system~\citep{willems2005note, markovsky2022identifiability}. Each is partitioned into a `past' block of $T_\mathrm{ini}$ steps, which, for a sufficiently long $T_\mathrm{ini}$, represents the initial condition of the prediction, and a `future' block of $N$ steps that represents the predicted trajectory $N$ steps in the future. Let us, for brevity, denote:
\begin{equation}
\begin{bmatrix} U_p \\ U_f \end{bmatrix} = H_L(u^d), \qquad \begin{bmatrix} Y_p \\ Y_f \end{bmatrix} = H_L(y^d).
\label{eq:hankel_uy}
\end{equation}
By Willems' fundamental lemma~\citep{willems2005note}, for an LTI system with persistently exciting input, every length-$L$ trajectory can be expressed as a linear combination of these columns. Equivalently, a sequence $(u, y)$ is a valid system trajectory if and only if, under sufficiently rich data, there exists a vector $g$ such that $H_L(u^d)\,g = u$ and $H_L(y^d)\,g = y$.
The Hankel matrices thus serve as a nonparametric predictor. Driven by the most recent $T_\mathrm{ini}$ measurements $(u_\mathrm{ini}, y_\mathrm{ini})$, the past block implicitly captures the system's current condition, playing the role of a state without requiring an explicit state estimate, while the future block generates the corresponding prediction. For a sufficiently large $T_\mathrm{ini}$, this past trajectory determines a consistent state, in the sense of~\citep{markovsky2021behavioral}.

This exact representation presumes a deterministic LTI system, which the system in~\eqref{eq:nl_statespace} is not, as it is noisy and nonlinear. We therefore adopt a regularized DeePC formulation~\citep{coulson2019data, verheijen2023handbook} to counteract noise and nonlinearity, in which a slack variable $\sigma_y$ relaxes the initial-output condition and a regularization term in the cost penalizes the vector $g$.
Over a prediction horizon of length $N$, the controller solves
{\setlength{\mathindent}{4pt}
\begin{subequations}\label{eq:deepc}
\begin{align}
\min_{g,\,u,\,y,\,\sigma_y}\quad & \ell(u, y) + \lambda_g \|g\|_2^2 + \lambda_{\sigma_y} \|\sigma_y\|_2^2 \label{eq:deepc_cost}\\
\text{s.t.}\quad & \begin{bmatrix} U_p \\ Y_p \\ U_f \\ Y_f \end{bmatrix} g = \begin{bmatrix} u_\mathrm{ini} \\ y_\mathrm{ini} \\ u \\ y \end{bmatrix} + \begin{bmatrix} 0 \\ \sigma_y \\ 0 \\ 0 \end{bmatrix}, \label{eq:deepc_data}\\
& u_k \in \mathcal{U}, \quad y_k \in \mathcal{Y}, \quad \forall k \in \mathbb{N}_{[0,\,N-1]}. \label{eq:deepc_constr}
\end{align}
\end{subequations}
}
Here $\ell(u, y)$ denotes the control cost accumulated over the prediction horizon, as introduced in~\eqref{eq:cost}, penalizing comfort violations and the control effort. The sets $\mathcal{U}$ and $\mathcal{Y}$ are the input and output constraint sets, and $\lambda_g, \lambda_{\sigma_y} > 0$ weight the regularization of $g$ and the slack against the control objective. As in receding-horizon control, only the first optimized input is applied, after which the horizon is shifted, and the problem is resolved in the next time step.
Regularization has also been shown experimentally to make the data-driven predictor robust to nonlinearities~\citep{dorfler2022bridging, elokda2021data}, though only to a limited extent~\citep{giacomelli2025insights, zieglmeier2025auv}.

\noindent \textbf{Incorporating Disturbances}\label{sec:deepc_dist}
\\
The formulation above describes the output as a function of the control input alone, yet the disturbance $d_k$ of~\eqref{eq:nl_statespace} acts on the room as well. Being available to the controller but not manipulable, it is included directly in the data, extending the input--output pairs to input--disturbance--output triplets $(u^d, d^d, y^d)$. A single vector $g$ continues to parameterize input, disturbance, and output as one trajectory. Constraining the future disturbance to its forecast $\hat{d}$ therefore conditions the predicted output on the anticipated external conditions, allowing the controller to account for them in advance. Therefore, a third Hankel matrix $[D_p^\top, D_f^\top]^\top$ is formed accordingly to~\eqref{eq:hankel_uy}. 
A measured disturbance can be interpreted as an additional, non-manipulable input, so that the predictor is now driven by the combined input--disturbance sequence, which must be persistently exciting. Since the disturbance cannot be shaped by design, this richness must instead be supplied by data spanning a representative range of operating and weather conditions.

The essential difference in handling the disturbance, compared to~\eqref{eq:deepc}, is that the future disturbance is \emph{not} a decision variable. Over the prediction horizon, it is fixed to a forecast $\hat{d}$, which lets the controller plan around influences it can anticipate but not command. Because one $g$ has to satisfy all blocks simultaneously, fixing the future disturbance also restricts the solution space for vector $g$, and with it the input trajectories the controller can realize. Since neither the noisy data nor an imperfect forecast can be reproduced exactly, this condition is relaxed with a slack variable $\sigma_d$. The disturbance-augmented problem then reads
{\setlength{\mathindent}{0pt}
\begin{subequations}\label{eq:deepc_dist}
\begin{align}
\min_{g,\,u,\,y,\,\sigma_y,\,\sigma_d}\quad & \ell(u, y) + \lambda_g \|g\|_2^2 + \lambda_{\sigma_y} \|\sigma_y\|_2^2 + \lambda_{\sigma_d} \|\sigma_d\|_2^2 \label{eq:deepc_dist_cost}\\
\text{s.t.}\quad & \begin{bmatrix} U_p \\ D_p \\ Y_p \\ U_f \\ D_f \\ Y_f \end{bmatrix} g = \begin{bmatrix} u_\mathrm{ini} \\ d_\mathrm{ini} \\ y_\mathrm{ini} \\ u \\ \hat{d} \\ y \end{bmatrix} + \begin{bmatrix} 0 \\ 0 \\ \sigma_y \\ 0 \\ \sigma_d \\ 0 \end{bmatrix}, \label{eq:deepc_dist_data}\\
& u_k \in \mathcal{U}, \quad y_k \in \mathcal{Y}, \quad \forall k \in \mathbb{N}_{[0,\,N-1]}. \label{eq:deepc_dist_constr}
\end{align}
\end{subequations}
}
The initial condition $(u_\mathrm{ini}, d_\mathrm{ini}, y_\mathrm{ini})$ is set to the recent measurements, while the future disturbance $\hat{d}$ is set to the forecast. The input is therefore optimized against the disturbances expected over the horizon.
The formulation~\eqref{eq:deepc_dist} is the common basis for all controllers compared in this work. The three nonlinear methods differ in how the data-driven predictor~\eqref{eq:deepc_dist_data} is formulated, which we detail in the following subsections. 

\subsection{Select-DPC}\label{sec:select-DPC}
DeePC builds its predictor from the entire dataset, spanning the full range of the system's behavior, and leaves the resulting operating point dependence to the regularization. 
Select-DPC instead retains, at each sampling instant, only the columns whose recorded conditions are `closest' to the current operating point, such that the selected columns span a neighborhood in which the dynamics are approximately linear~\citep{beerwerth2025morecontextualsamplingnonlinear, naf2025choose}. 
The general idea is that the local linearization lets a linear combination predict well in the neighborhood while the optimization remains convex, at the cost of a data-selection step at each sampling instant. 
The scheduled approaches that follow localize on the scheduling variable $\rho_k$, the sun's position, alone. The selection here instead draws on the room's full recent input--disturbance--output history $(u_\mathrm{ini}, d_\mathrm{ini}, y_\mathrm{ini})$, matching columns on the complete operating condition rather than on $\rho_k$ only.

Let the full data matrices in~\eqref{eq:deepc_dist} have $M = T - L + 1$ columns, indexed by $i \in \mathbb{N}_{[1,\,M]}$, and let $(\cdot)_{:,\mathcal{I}}$ denote the submatrix formed by the columns in an index set $\mathcal{I}$.
At each discrete time step $k$, Select-DPC selects a subset $\mathcal{I}_k$ of $N_s \ll M$ columns and assembles the reduced blocks $U_p^{(k)} = (U_p)_{:,\mathcal{I}_k}$, and analogously $D_p^{(k)}, Y_p^{(k)}, U_f^{(k)}, D_f^{(k)}, Y_f^{(k)}$. 
Since the selection is repeated at every step, these matrices, and hence the predictor itself, are time-varying, which the index $(k)$ makes explicit. Restricting the predictor to data from a neighborhood of the current operating point yields a local linear approximation of the nonlinear dynamics. 

The relevance of each column is quantified by its closeness to the current operating point, for which several metrics have been proposed, including norm-based distances in trajectory space, distances in a learned low-dimensional embedding, and metrics that additionally account for the future reference~\citep{beerwerth2025morecontextualsamplingnonlinear, naf2025choose}. 
For the heating application, this current operating point is the room's recent thermal state, comprising the heating power $u$, the disturbances $d$, and the resulting temperature $y$ over the last $T_\mathrm{ini}$ steps. As the simplest norm-based distance, we adopt the Euclidean distance between the past input--disturbance--output window of each column and the most recent measured triplet,
\begin{equation}
\delta_i^{(k)} = \left\| \begin{bmatrix} U_{p,i} \\ D_{p,i} \\ Y_{p,i} \end{bmatrix} - \begin{bmatrix} u_\mathrm{ini} \\ d_\mathrm{ini} \\ y_\mathrm{ini} \end{bmatrix} \right\|_2, \qquad i \in \mathbb{N}_{[1,\,M]}, \label{eq:select_dist}
\end{equation}
where $U_{p,i}$ denotes the $i$-th column of $U_p$, and the index set $\mathcal{I}_k$ retains the $N_s$ closest columns,
\begin{equation}
\mathcal{I}_k = \operatorname*{arg\,min}_{|\mathcal{I}| = N_s} \sum_{i \in \mathcal{I}} \delta_i^{(k)}. \label{eq:select_set}
\end{equation}
%
% New:
\rev{The distance is computed on the signals in their physical units without normalization. This keeps the selection consistent with the subsequent optimization, in which the relaxed conditions of the data equation are penalized quadratically in the same units, so that both prioritize the same channels. It also keeps Select-DPC directly comparable to the other predictive controllers, which form the linear combination on the same unscaled data.}
Select-DPC then solves the disturbance-augmented problem~\eqref{eq:deepc_dist} at each time step with the full data matrices in~\eqref{eq:deepc_dist_data} replaced by their selected, time-varying counterparts 
{\setlength{\mathindent}{0pt}
\begin{subequations}\label{eq:S_deepc_dist}
\begin{align}
\min_{g,\,u,\,y,\,\sigma_y,\,\sigma_d}\quad & \ell(u, y)\!+\!\lambda_g \|g\|_2^2\!+\!\lambda_{\sigma_y} \|\sigma_y\|_2^2\!+\!\lambda_{\sigma_d} \|\sigma_d\|_2^2 \label{eq:S_deepc_dist_cost}\\
\text{s.t.}\quad & \begin{bmatrix} U_p^{(k)} \\ D_p^{(k)} \\ Y_p^{(k)} \\ U_f^{(k)} \\ D_f^{(k)} \\ Y_f^{(k)} \end{bmatrix} g = \begin{bmatrix} \vphantom{U_p^{(k)}} u_\mathrm{ini} \\ \vphantom{D_p^{(k)}} d_\mathrm{ini} \\ \vphantom{Y_p^{(k)}} y_\mathrm{ini} \\ \vphantom{U_f^{(k)}} u \\ \vphantom{D_f^{(k)}} \hat{d} \\ \vphantom{Y_f^{(k)}} y \end{bmatrix} + \begin{bmatrix} \vphantom{U_p^{(k)}} 0 \\\vphantom{D_p^{(k)}} 0 \\ \vphantom{Y_p^{(k)}} \sigma_y \\ \vphantom{U_f^{(k)}} 0 \\ \vphantom{D_f^{(k)}} \sigma_d \\ \vphantom{Y_f^{(k)}} 0 \end{bmatrix}, \label{eq:S_deepc_dist_data}\\
& u_k \in \mathcal{U}, \quad y_k \in \mathcal{Y}, \quad \forall k \in \mathbb{N}_{[0,\,N-1]}. \label{eq:S_deepc_dist_constr}
\end{align}
\end{subequations}}%
and $g \in \mathbb{R}^{N_s}$. The cost, regularization, slack variables, and constraint sets are identical to those of DeePC, while only the data entering the predictor changes from one step to the next.

\subsection{Gain-Scheduling DPC}\label{sec:GS-DPC}
GS-DPC addresses nonlinear behavior by formulating the classical idea of gain scheduling~\citep{rugh2000research} in a data-driven setting. Instead of one global predictor, it maintains a family of local data representations, each associated with a region of the operating range. Based on the current measurement of the scheduling variable, which indicates the current operating range, the corresponding local data representation is used in the prediction~\citep{guerrero2025gain, zieglmeier2025gain}.
Within a sufficiently narrow region, the dynamics are approximately linear, i.e., the corresponding regional data acts as a local linear approximation of the nonlinear system.

GS-DPC uses the solar azimuth $\rho_k$, defined in Section~\ref{sec:methods}, directly as its scheduling variable, partitioning its range into $n$ regions $\{M_r\}_{r=1}^{n}$, with region $r$ covering the interval between the boundaries $[b_r,\, b_{r+1}]$. The daytime arc, over which the response to solar radiation varies, is split into $15^\circ$ windows, from $15^\circ$ to $300^\circ$, giving nineteen daytime regions. The remaining sector from $300^\circ$ through $0^\circ$ to $15^\circ$, over which the sun is down and provides no solar gain, is collected into a single overnight region. This yields $n = 20$ regions in total, each grouping the trajectories that were recorded while the sun was within that angular sector. Assigning every sample to exactly one region corresponds to representing the azimuth as a binary vector with a single nonzero entry, a one-hot encoding, the same representation used for the input features in~\citep{bunning2022physics}.

The data for region $r$ are collected from the recorded input--disturbance--output trajectory wherever $\rho_k$ lies within the boundaries $[b_r,\, b_{r+1}]$. At inference, when $\rho_k$ enters region $r_k$ at step $k$, the predictor initializes on the $T_\mathrm{ini}$ most recent samples and predicts over the next $N$ steps. A region's data would only cover the time $\rho_k$ actually spends inside $[b_r,\, b_{r+1}]$. But the predictor still needs the $T_\mathrm{ini}$ steps before entry to initialize, and the $N$ steps after exit to predict, and these steps lie outside the region. The offline data for the region must cover the same span the predictor needs online, so each in-region trajectory is extended by $T_\mathrm{ini}$ samples before its entry index and by $N$ samples after its exit index. Each time $\rho_k$ stays continuously within $[b_r,\, b_{r+1}]$, from an entry index $\underline{k}$ to an exit index $\overline{k}$, the recorded data over that extended segment is
\begin{equation}
\big[\,\underline{k} - T_\mathrm{ini},\ \ \overline{k} + N\,\big]. \label{eq:gs_extension}
\end{equation}
The trajectories within these extended segments are partitioned into past and future blocks as in~\eqref{eq:hankel_uy}. Since the azimuth evolves periodically, each region is entered once per day, so the recorded data yields one extended segment per region and day. The corresponding Hankel matrices are concatenated column-wise into a single regional mosaic-Hankel matrix~\citep{van2020willems}, which forms the regional blocks $U_p^{(r_k)}, D_p^{(r_k)}, Y_p^{(r_k)}, U_f^{(r_k)}, D_f^{(r_k)}, Y_f^{(r_k)}$. The active region, and therefore the regional blocks, are held fixed over the prediction horizon.

The active region is selected online by assigning the current scheduling variable to its interval,
\begin{equation}
r_k = \{\, r : \rho_k \in [b_r,\, b_{r+1}] \,\}, \label{eq:gs_select}
\end{equation}
and the corresponding regional blocks are loaded into the predictor. Because $\rho_k$ evolves smoothly with the time of day, the regions are switched directly, without additional switching logic, cf.~\citep{zieglmeier2025gain}.

GS-DPC then solves the disturbance-augmented problem~\eqref{eq:deepc_dist} at each step with the full data matrices replaced by those of the active region $r_k$ in~\eqref{eq:GS_deepc_dist}.
{\setlength{\mathindent}{0pt}
\begin{subequations}\label{eq:GS_deepc_dist}
\begin{align}
\min_{g,\,u,\,y,\,\sigma_y,\,\sigma_d}\quad & \ell(u, y)\!+\!\lambda_g \|g\|_2^2\!+\!\lambda_{\sigma_y} \|\sigma_y\|_2^2\!+\!\lambda_{\sigma_d} \|\sigma_d\|_2^2 \label{eq:GS_deepc_dist_cost}\\
\text{s.t.}\quad & \begin{bmatrix} U_p^{(r_k)} \\ D_p^{(r_k)} \\ Y_p^{(r_k)} \\ U_f^{(r_k)} \\ D_f^{(r_k)} \\ Y_f^{(r_k)} \end{bmatrix} g = \begin{bmatrix} \vphantom{U_p^{(r_k)}} u_\mathrm{ini} \\ \vphantom{D_p^{(r_k)}} d_\mathrm{ini} \\ \vphantom{Y_p^{(r_k)}} y_\mathrm{ini} \\ \vphantom{U_f^{(r_k)}} u \\ \vphantom{D_f^{(r_k)}} \hat{d} \\ \vphantom{Y_f^{(r_k)}} y \end{bmatrix} + \begin{bmatrix} \vphantom{U_p^{(r_k)}} 0 \\ \vphantom{D_p^{(r_k)}} 0 \\ \vphantom{Y_p^{(r_k)}} \sigma_y \\ \vphantom{U_f^{(r_k)}} 0 \\ \vphantom{D_f^{(r_k)}} \sigma_d \\ \vphantom{Y_f^{(r_k)}} 0 \end{bmatrix} \label{eq:GS_deepc_dist_data}\\
& u_k \in \mathcal{U}, \quad y_k \in \mathcal{Y}, \quad \forall k \in \mathbb{N}_{[0,\,N-1]} \label{eq:GS_deepc_dist_constr}
\end{align}
\end{subequations}}%
Since switching only substitutes the data matrices, the cost, regularization, slack variables, constraints, and per-step optimization are identical to those of DeePC. 
The distinction is that each predictor is built from shorter, locally informative data rather than from one large global dataset, which improves local prediction accuracy while reducing the size of the matrices entering the optimization.

\subsection{Linear Parameter-Varying DPC}\label{sec:lvp-dpc}
LPV-DPC retains the global data set and incorporates the scheduling dependence into the predictor itself~\citep{verhoek2021data, verhoek2025linear}, such that the image spanned by the data matrices is varying based on the (measured and predicted) solar azimuth.
Because the LPV system is linear along a given scheduling trajectory, the predictor retains a linear structure whose dynamics track the azimuth as it evolves over the horizon, rather than being frozen at its current value.
LPV-DPC extends DeePC to the class of LPV systems and thereby to nonlinear systems that admit an LPV embedding with a measurable scheduling variable, for which an LPV extension of the fundamental lemma holds~\citep{verhoek2025behavioural, verhoek2021fundamental}.

For the offline trajectory, define the scheduling-weighted sequences
\begin{equation}
    u^{\rho,d}_k := \rho^d_k \otimes u^d_k,
\end{equation}
and analogously $d^{\rho,d}_k$ and $y^{\rho,d}_k$, where $\rho^d_k$ is the scheduling variable recorded alongside the data.
To keep the scheduling-weighted data computational well scaled, the azimuth is normalized to the unit interval, $\rho_k = \mathrm{az}_k / 360^\circ \in [0,1]$.
Their depth-$L$ Hankel matrices are partitioned into past and future blocks exactly as in~\eqref{eq:hankel_uy}, yielding $U_p^{\rho}, U_f^{\rho}, D_p^{\rho}, D_f^{\rho}, Y_p^{\rho}, Y_f^{\rho}$.
Treating these scheduling-weighted samples as auxiliary free variables, as if they were additional inputs, embeds the LPV behavior into that of a higher-dimensional LTI system, to which the data-driven representation of Willems' fundamental lemma applies. Since the auxiliary variables are in fact not free, this LTI embedding over-approximates the LPV behavior. The scheduling structure is recovered by constraining the auxiliary variables to equal the actual scheduling-weighted products $\rho_k \otimes u_k$, $\rho_k \otimes d_k$, and $\rho_k \otimes y_k$, a restriction that eliminates the over-approximation introduced by the embedding~\citep{verhoek2025behavioural}.

Let $\Pi_p$ and $\Pi_f$ be the block-diagonal weighting matrices whose $k$-th diagonal block is $\rho_k \otimes I$, with $I$ of appropriate dimension.
The blocks of $\Pi_p$ are assembled from the measured past azimuth window $\rho_\mathrm{ini}$ over the last $T_\mathrm{ini}$ steps, while those of $\Pi_f$ are taken from the azimuth sequence $\bar{\rho}_{[0,\,N-1]}$ over the prediction horizon.
The azimuth is exogenous, fixed by the building's location and the time and date, so $\bar{\rho}_{[0,\,N-1]}$ is available exactly in advance and $\Pi_f$ is known at solve time. 
Constraining each scheduling-weighted block to equal its weighted counterpart, $U_p^{\rho}g = \Pi_p U_pg$, and likewise for the remaining blocks, and moving these to one side, appends to the data equation~\eqref{eq:deepc_dist_data} the scheduling-consistency rows
\begin{equation}
\begin{bmatrix} U_p^{\rho} - \Pi_p U_p \\ D_p^{\rho} - \Pi_p D_p \\ Y_p^{\rho} - \Pi_p Y_p \\ U_f^{\rho} - \Pi_f U_f \\ D_f^{\rho} - \Pi_f D_f \\ Y_f^{\rho} - \Pi_f Y_f \end{bmatrix} g = \begin{bmatrix} 0 \\ 0 \\ \sigma_y^\rho \\ 0 \\ \sigma_d^\rho \\ 0 \end{bmatrix}. \label{eq:lpv_data}
\end{equation}
LPV-DPC then solves at each step
\begin{subequations}\label{eq:lpv}
\begin{align}
\min_{\substack{g,\,u,\,y,\\ \sigma_y,\,\sigma_d,\,\sigma_y^\rho,\,\sigma_d^\rho}}\quad & \text{\eqref{eq:deepc_dist_cost}} + \lambda_{\sigma_y^\rho} \|\sigma_y^\rho\|_2^2 + \lambda_{\sigma_d^\rho} \|\sigma_d^\rho\|_2^2 \label{eq:lpv_cost}\\
\text{s.t.}\quad & \eqref{eq:deepc_dist_data},\ \eqref{eq:lpv_data},\ \eqref{eq:deepc_dist_constr}. \label{eq:lpv_constr}
\end{align}
\end{subequations}
The two scheduling-consistency slacks $\sigma_y^\rho$ and $\sigma_d^\rho$ are weighted by $\lambda_{\sigma_y^\rho}, \lambda_{\sigma_d^\rho} > 0$.

Throughout, the global data matrices are used. Relative to DeePC, the decision variable $g$ keeps the column dimension of the full data set, and the optimization grows only by the scheduling-consistency rows~\eqref{eq:lpv_data} and the two slack variables that relax them.
While Select-DPC and GS-DPC approximate the nonlinear dynamics by local linearity and carry no representation guarantee, the scheduling-weighted data equation provides an exact data-driven representation for LPV systems with shifted-affine scheduling dependence, provided the data satisfies a generalized persistency-of-excitation condition on the recorded signals and their scheduling-weighted products~\citep{verhoek2025behavioural}.
For the heating system, this exactness holds to the extent that the dynamics admit such an LPV embedding in the azimuth. This gives LPV-DPC the strongest theoretical foundation among the methods considered here.

%% -------------------------------------------------------
%% SECTION 5: SIMULATION RESULTS
%% -------------------------------------------------------
\section{Simulation and Comparison}\label{sec:sim-results}
The four data-driven controllers and the hysteresis benchmark are evaluated on the digital twin of the UMAR unit over a full heating season. Throughout this section, the controllers receive the measured disturbances over the prediction horizon, so the forecast is exact. Forecast accuracy would otherwise limit the achievable closed-loop performance, and the results would reflect the quality of the forecast more than the control formulations under comparison. The following subsections describe the generation of the data, the closed-loop results, the tightening of the comfort band, and the hyperparameters used.

\subsection{Data Generation}\label{sec:data-generation}
All data-driven controllers are constructed from offline input--disturbance--output trajectories generated on the digital twin, driven throughout by the on-site measurements of the recorded season. Only the room-temperature response is simulated, which lets the excitation be designed freely, and the experiment be repeated under identical conditions.
For the recorded data to be sufficiently rich to identify the predictor, each room is driven in open loop with a randomized heating input, the valve-open fraction at each interval drawn independently from a beta distribution as $\tau_j \sim \mathrm{Beta}(1, 7)$.
The excitation spans the full input range, with a low mean to keep the unit from overheating during open-loop data collection. Unlike a uniform or Gaussian distribution, whose mean is fixed at the midpoint of its support, the right-skewed beta distribution combines full-range support with an independently tunable low mean.
% New:
\rev{The resulting input Hankel matrix $H_L(u^d)$ has full row rank. For the disturbance-augmented matrix, a rank test is not informative, since measurement noise renders it full rank irrespective of its information content, and the disturbance cannot be designed. Its richness is instead ensured by the data set spanning the full outdoor-temperature and solar range of the heating season.}

Over a long horizon, even this low-mean forcing lets the room temperature drift slowly out of range during warm or sunny weeks, so the data is generated as a set of independent one-week batches rather than one continuous trajectory. 
Each batch is a separate simulation starting from the same settled initial thermal state, so the slow drift does not accumulate across batches while each batch still spans a full week of weather and day--night cycling. 
The batches are combined into a single mosaic-Hankel matrix~\citep{van2020willems}, so that no column spans a batch boundary and the discontinuity between batches never enters a prediction.

To keep data collection and evaluation clearly separated, the data matrices are built from the winter 2019--2020 period of $26$ weekly batches at a $15$-minute control resolution, and the controllers are evaluated in closed loop over the winter 2020--2021, from \formatdate{17}{10}{2020} to \formatdate{28}{2}{2021}.
GS-DPC and Select-DPC use the full data set, since both reduce the active data matrices online by regional selection and by column selection, respectively. 
Standard DeePC and LPV-DPC solve with a single fixed set of data matrices. As the computational complexity grows with the number of columns~\citep{berberich2020data}, using the full data set for these two approaches is too expensive. These therefore use the four specific batches of Table~\ref{tab:batches}, the warmest, coldest, darkest, and sunniest weeks, which span the outdoor-temperature and solar ranges of the data-collection period.
% New: 
\rev{At each step, Select-DPC and GS-DPC solve with at most $1521$ columns, fewer than the $2628$ columns of DeePC and LPV-DPC (Table~\ref{tab:hyperparameters}), so their advantage follows from which columns enter the predictor rather than from the amount of data.}
\begin{table}[t]
\centering
\caption{The four batches used by standard DeePC and LPV-DPC, with the mean outdoor temperature $\overline{T}_\mathrm{out}$ and the cumulative solar irradiation $E_\mathrm{sol}$ over the week.}
\label{tab:batches}
\begin{tabular}{llcc}
\toprule
Week (start) & Extreme & $\overline{T}_\mathrm{out}$ [$^\circ$C] & $E_\mathrm{sol}$ [kWh/m$^2$] \\
\midrule
\formatdate{17}{10}{2019} & warmest & $13.7$ & $11.3$ \\
\formatdate{19}{12}{2019} & darkest & $5.7$  & $3.7$  \\
\formatdate{26}{12}{2019} & coldest & $1.9$  & $4.4$  \\
\formatdate{20}{2}{2020}  & sunniest & $8.6$ & $17.2$ \\
\bottomrule
\end{tabular}
\end{table}

\subsection{Simulation Results}
\label{subsec:results}
The comparison covers the full evaluation winter, a representative week, and four extreme weeks to cover the full seasonal operating range. Two evaluation metrics are reported throughout: the heating energy $E$ and the comfort-band violation $V = \sum_k (\underline{s}_k + \overline{s}_k)\,\Delta k$, the accumulated violation~\eqref{eq:band_slack} weighted by its duration, reported in Kelvin-hours (K\,h). Each of the three rooms is regulated by its own independent controller, % and the reported energy is the actual consumption $E=\sum_k u_k\,\Delta k$. 
% Replaced by: 
\rev{and the reported energy is the delivered heating energy $E=\sum_k P_{\mathrm{h},k}\,\Delta k$. The delivered power exceeds the commanded power, since the pulse-width modulation of the inner loop can realize valve openings only in full minutes. Averaged over all active control intervals, this surplus amounts to about $3\%$ of the maximum power of each room, as detailed in Appendix~\ref{app:power}.}
Although the metrics are reported per room, the evaluation is based on the summed total of all rooms, since the rooms are thermally coupled and one can be improved at the expense of another. Throughout, three comparisons structure the discussion: the data-driven controllers against the hysteresis benchmark, DeePC against the three nonlinear methods, and the three nonlinear methods against one another. Hyperparameters were tuned separately for each controller and room by grid search, detailed in Section~\ref{sec:hp-tuning}.

\subsubsection{Representative Week}
\begin{figure*}[t]
  \centering
  \includegraphics[width=0.9\textwidth]{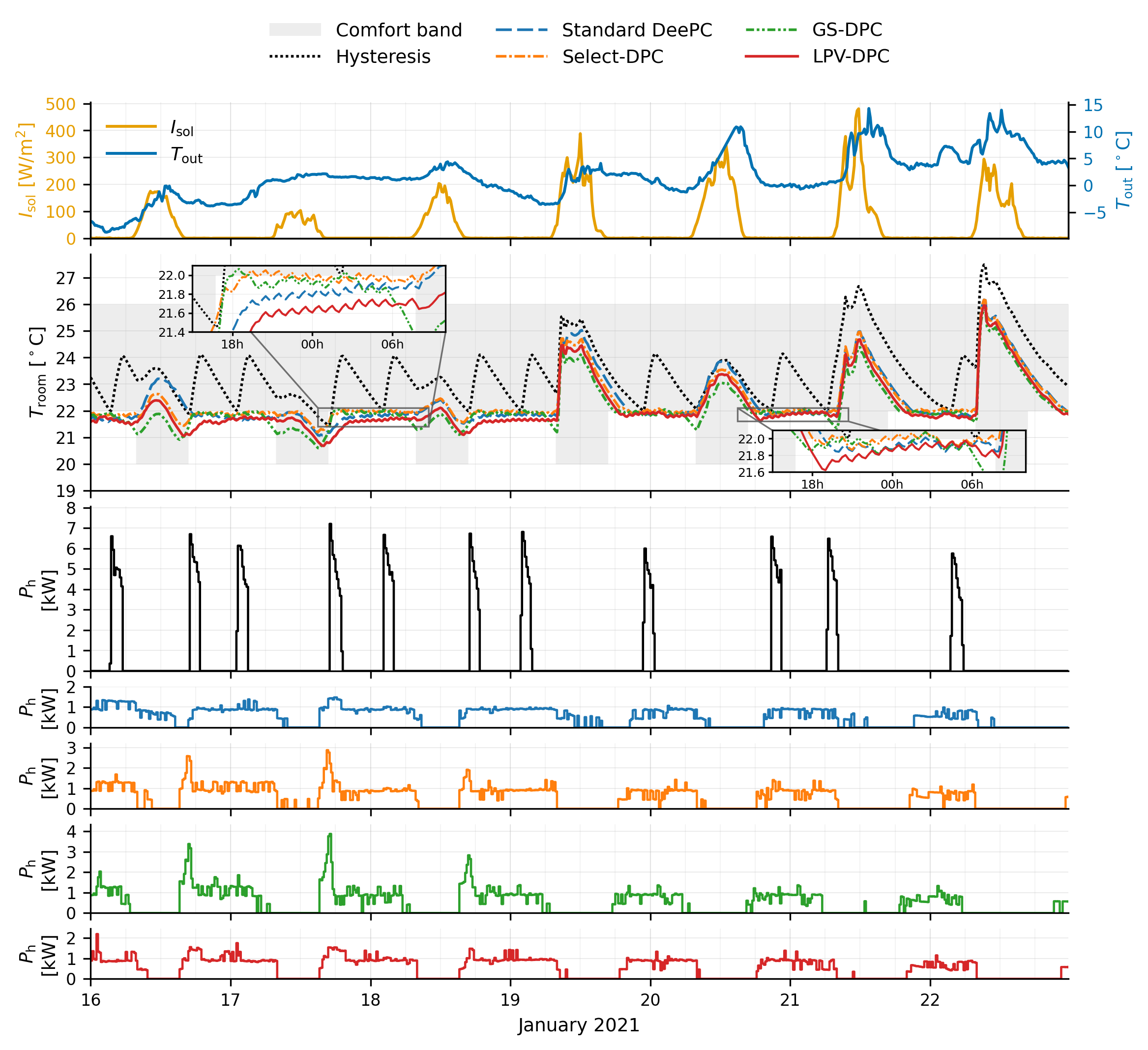}
    \caption{Closed-loop behavior of the five controllers over the representative week (16--22 January 2021) in room R273. First: measured solar irradiance and outdoor temperature. Second: room temperature with the time-varying comfort band shaded, with two insets detailing the tracking near the lower bound. Bottom five: heating power of each controller in the order: hysteresis control, DeePC, Select-DPC, GS-DPC, LPV-DPC.}
    \label{fig:R273_Week13}
\end{figure*}

The representative week (Table~\ref{tab:wk-average}, illustrated for room R273 in Figure~\ref{fig:R273_Week13}) shows the controllers under typical mid-winter conditions in Switzerland. It is representative in that it spans the full spread of environmental conditions within a single week, from cold to mild outdoor temperatures and from overcast to sunny days. 
In terms of energy, all controllers lie within a narrow band around $170$--$180$\,kWh, with GS-DPC and Select-DPC the lowest and hysteresis and standard DeePC the highest. The nonlinear methods therefore undercut the hysteresis controller by up to $7.3\%$.
In terms of comfort, hysteresis and Select-DPC keep the violations smallest, whereas LPV-DPC clearly violates the band the most.
\begin{table}[t]
\centering
\footnotesize
\setlength{\tabcolsep}{3pt}
\caption{Heating energy and comfort-band violation per controller over the representative week (16--22 January 2021), per controlled room and summed over all three rooms.}
\label{tab:wk-average}
\begin{tabular}{l rrrr rrrr}
\toprule
& \multicolumn{4}{c}{$E$ [kWh]} & \multicolumn{4}{c}{$V$ [K\,h]} \\
\cmidrule(lr){2-5}\cmidrule(lr){6-9}
Controller & R272 & R273 & R274 & Total & R272 & R273 & R274 & Total \\
\midrule
Hysteresis & 44.7 & 102.7 & 34.2 & 181.5 & 0.2 & 4.8 & 0.0 & 5.0 \\
DeePC & 48.9 & 88.4 & 43.1 & 180.3 & 4.7 & 15.9 & 0.0 & 20.6 \\
Select-DPC & 46.0 & 93.9 & 35.6 & 175.5 & 6.6 & 5.2 & 3.2 & 15.0 \\
GS-DPC & 45.6 & 86.5 & 36.1 & 168.2 & 4.6 & 14.2 & 2.2 & 21.0 \\
LPV-DPC & 45.4 & 90.3 & 36.0 & 171.7 & 26.1 & 28.1 & 9.9 & 64.1 \\
\bottomrule
\end{tabular}
\end{table}

The behavior visible in Figure~\ref{fig:R273_Week13} reflects this: the predictive controllers exploit the forecast to preheat ahead of cold intervals and to withhold heat ahead of solar gains, and operate close to the lower comfort bound to save energy.
Figure~\ref{fig:R273_Week13} makes the effect of the external disturbances on the closed-loop directly visible. On the sunnier days (\formatdate{19}{1}{2021} onward), solar radiation entering the window raises the room temperature sharply from around 08:00 until roughly midday, after which the temperature falls again even while the measured irradiance remains high, since the sun no longer penetrates the window at that geometry. This is the operating point-dependent solar gain discussed in Section~\ref{sec:application}. 
The predictive controllers respond by withholding heat over the sunny interval, and the scheduled methods, GS-DPC in particular, reduce heating already before the radiation sets in, trading a small comfort-band violation for the avoided energy. The hysteresis controller has no such anticipation and, reacting only to the band edges, occasionally continues heating into a period when the solar gain is about to warm the room. This produces both a comfort violation, as the room is driven outside the upper bound of the comfort band, and wasted energy, since the free solar heat is not exploited.

The colder and darker days (before \formatdate{19}{1}{2021}) show the effect of the time-varying comfort band more clearly. When the lower bound is relaxed during the daytime occupancy schedule, the data-driven controllers exploit the widened band to suspend heating and let the room cool, and then preheat toward the evening so that the temperature returns within the tightened band by the time the room is occupied. The predictive controllers thus handle a time-varying comfort specification directly, allowing a relaxed band during absence to save energy while restoring comfort ahead of occupancy, which the reactive hysteresis controller cannot schedule.

The input pattern separates the reactive and predictive controllers most clearly. The hysteresis controller operates bang-bang, delivering full power or none, whereas the data-driven controllers apply lower instantaneous power spread over longer intervals and consequently hold the room temperature closer to the lower comfort bound. 
The comfort-band violations of the data-driven controllers are clearly visible where the room temperature is held constantly, slightly below the lower comfort bound. These violations follow directly from penalizing the band violation in the cost rather than enforcing it as a hard constraint.

The methods also differ in how consistently they track the band across the week. On the colder and darker days, LPV-DPC shows the largest violations below the band, with improved tracking of the lower bound on the milder later days. GS-DPC and DeePC exhibit the same tendencies, only weaker. Select-DPC tracks the band most consistently across both weather regimes. This is consistent with the selection mechanism: by drawing the columns closest to the current operating point into the predictor at every step, Select-DPC adapts its data to the prevailing conditions, whereas the other methods predict from a fixed set of data matrices that must serve both the cold and the mild parts of the week.

\subsubsection{Full Evaluation Winter}\label{sec:full-winter}
Over the full winter (Table~\ref{tab:winter-energy-comfort}), every data-driven controller consumes less energy than the hysteresis benchmark, with Select-DPC saving roughly $11\%$. All three nonlinear methods consume less energy than standard DeePC, with Select-DPC and GS-DPC the most economical and LPV-DPC close behind. Localizing the data to the operating point therefore pays off in energy over a single global representation.
\begin{table}[h!]
\centering
\footnotesize
\setlength{\tabcolsep}{3pt}
\caption{Heating energy and comfort-band violation per controller over the full evaluation winter (17 October 2020--28 February 2021), per controlled room and summed over all three rooms.}
\label{tab:winter-energy-comfort}
\begin{tabular}{l rrrr rrrr}
\toprule
& \multicolumn{4}{c}{$E$ [kWh]} & \multicolumn{4}{c}{$V$ [K\,h]} \\
\cmidrule(lr){2-5}\cmidrule(lr){6-9}
Controller & R272 & R273 & R274 & Total & R272 & R273 & R274 & Total \\
\midrule
Hysteresis & 772 & 1577 & 523 & 2872 & 38 & 192 & 29 & 259 \\
DeePC & 760 & 1234 & 695 & 2690 & 237 & 396 & 259 & 891 \\
Select-DPC & 701 & 1359 & 491 & 2552 & 104 & 192 & 58 & 353 \\
GS-DPC & 735 & 1320 & 509 & 2564 & 118 & 340 & 63 & 521 \\
LPV-DPC & 703 & 1367 & 535 & 2604 & 481 & 679 & 222 & 1382 \\
\bottomrule
\end{tabular}
\end{table}
On comfort, the ranking differs and is shaped by the same consequence of using the comfort band as a soft constraint within the cost function. Because the band enters the cost rather than the constraints, every predictive controller trades comfort against energy and operates near the lower bound, accepting occasional violations in exchange for saving energy. Select-DPC and GS-DPC keep the seasonal violation low and well below DeePC, so both localized methods improve on the global predictor in comfort as well as energy. 
LPV-DPC is the exception, with the highest violation of all controllers. 
Its theoretical advantage does not carry over here because measurement noise in the recorded data enters the scheduling-consistency relations~\eqref{eq:lpv_data} and degrades the accuracy of the scheduled prediction, an effect reported for parameter-varying data-driven predictors on noisy data~\citep[Chapters 7 and 8]{Verhoek_Thesis}. Relative to these predictive controllers, the hysteresis controller attains a lower seasonal violation.
% New:
\rev{A comparison of the energy consumption at comparable comfort follows in Section~\ref{sec:tightening}.}

\subsubsection{Extreme Weeks}
\begin{figure}
    \centering
    \begin{subfigure}[t]{\linewidth}
        \centering
        \includegraphics[width=\linewidth]{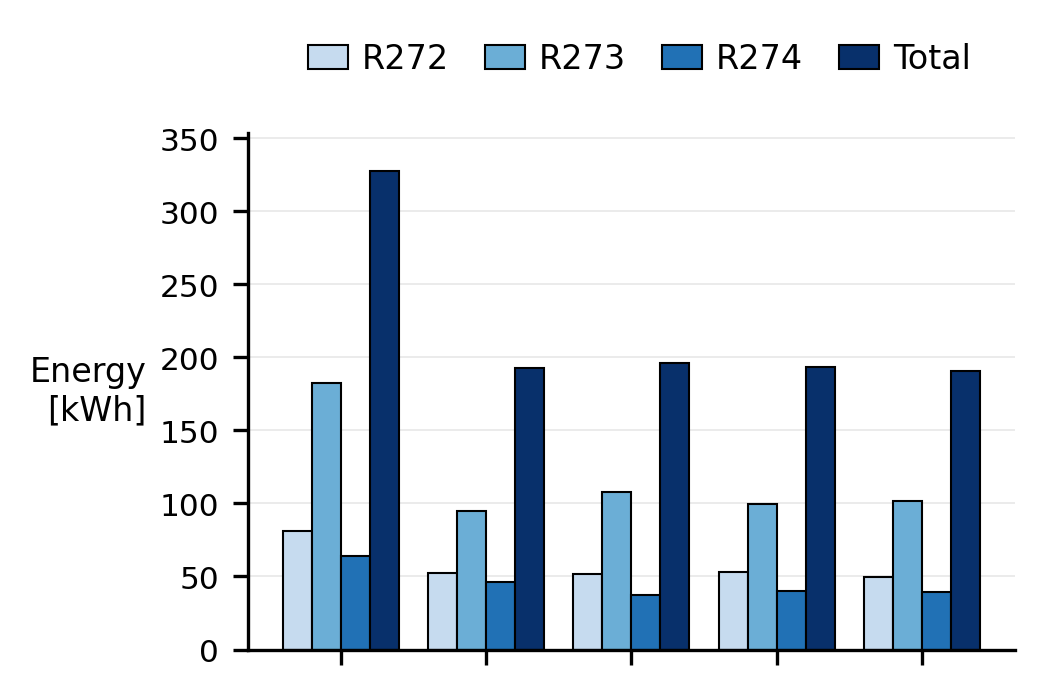}
        \caption{Coldest week}
        \label{fig:extreme-energy-coldest}
    \end{subfigure}

    \begin{subfigure}[t]{\linewidth}
        \centering
        \includegraphics[width=\linewidth]{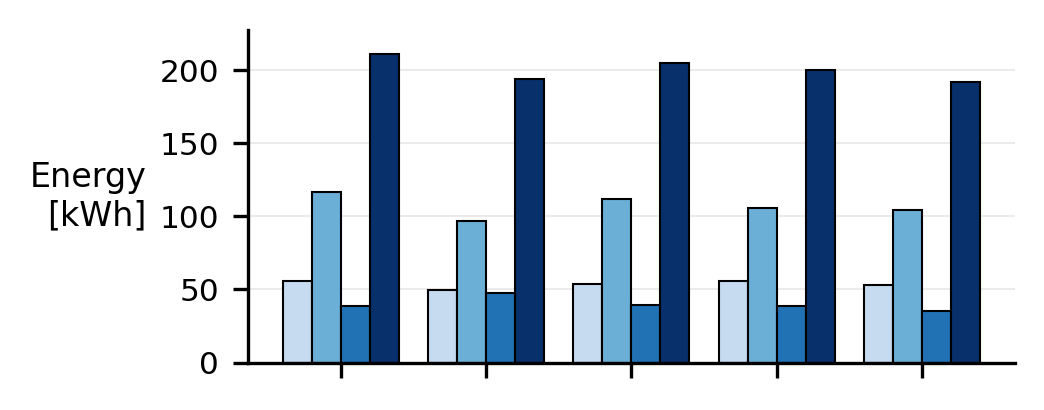}
        \caption{Darkest week}
        \label{fig:extreme-energy-darkest}
    \end{subfigure}

    \begin{subfigure}[t]{\linewidth}
        \centering
        \includegraphics[width=\linewidth]{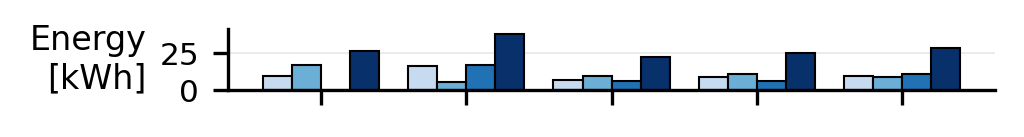}
        \caption{Warmest week}
        \label{fig:extreme-energy-warmest}
    \end{subfigure}

    \begin{subfigure}[t]{\linewidth}
        \centering
        \includegraphics[width=\linewidth]{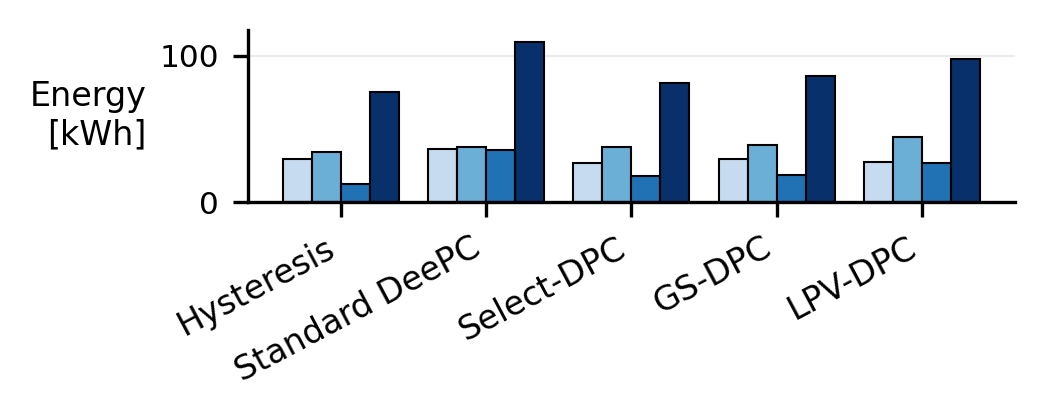}
        \caption{Sunniest week}
        \label{fig:extreme-energy-sunniest}
    \end{subfigure}
    \caption{Heating energy per controller and room across the four corner weeks, coldest (9--15 January 2021), darkest (28 November--4 December 2020), warmest (17--23 October 2020), and sunniest (13--19 February 2021).}
    \label{fig:extreme-energy-weeks}
\end{figure}

\begin{figure}
    \centering
    \begin{subfigure}[t]{\linewidth}
        \centering
        \includegraphics[width=\linewidth]{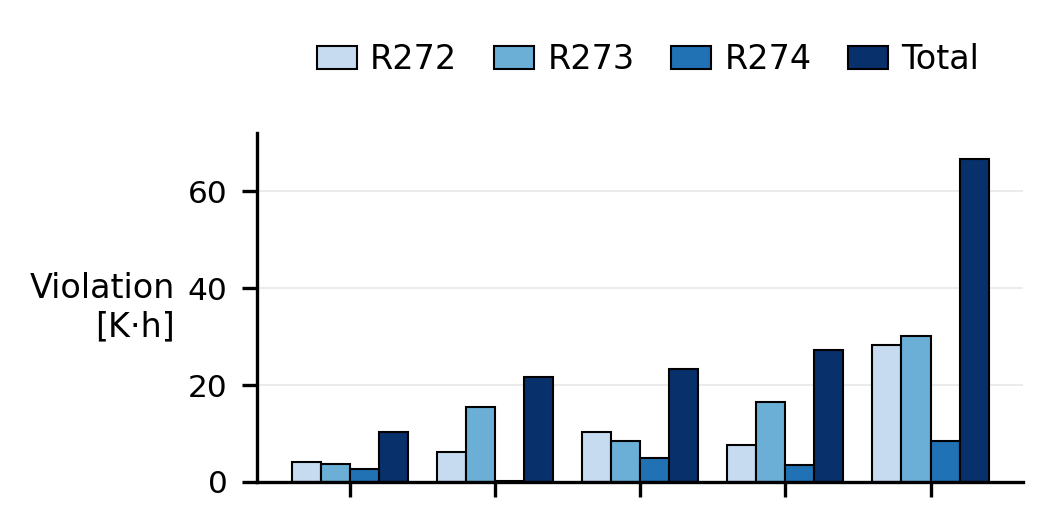}
        \caption{Coldest week}
        \label{fig:extreme-violation-coldest}
    \end{subfigure}

    \begin{subfigure}[t]{\linewidth}
        \centering
        \includegraphics[width=\linewidth]{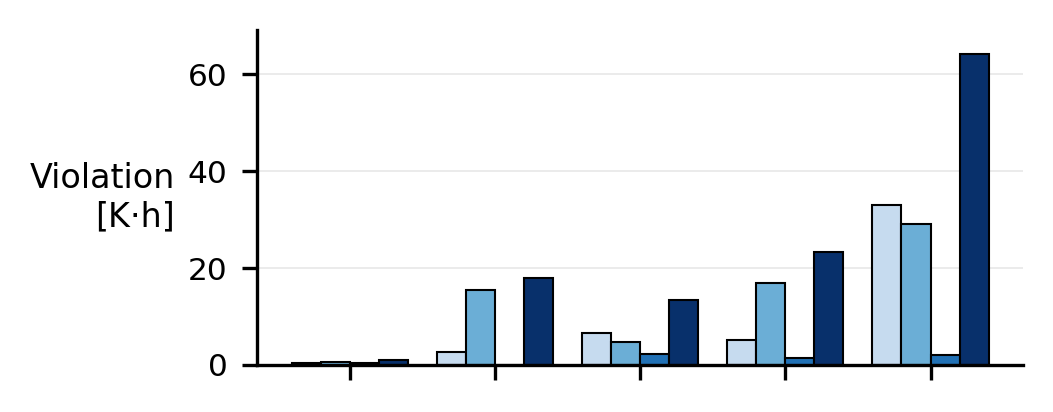}
        \caption{Darkest week}
        \label{fig:extreme-violation-darkest}
    \end{subfigure}

    \begin{subfigure}[t]{\linewidth}
        \centering
        \includegraphics[width=\linewidth]{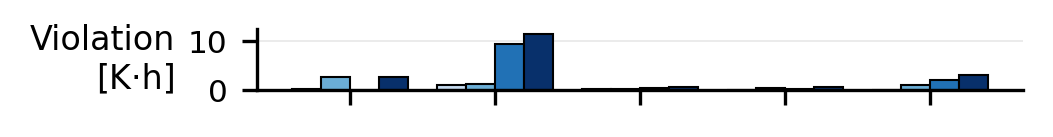}
        \caption{Warmest week}
        \label{fig:extreme-violation-warmest}
    \end{subfigure}

    \begin{subfigure}[t]{\linewidth}
        \centering
        \includegraphics[width=\linewidth]{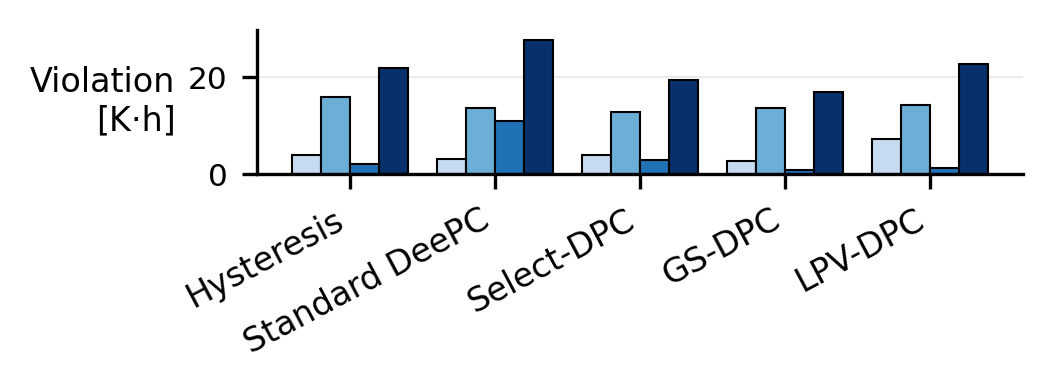}
        \caption{Sunniest week}
        \label{fig:extreme-violation-sunniest}
    \end{subfigure}
    \caption{Comfort violation per controller and room across the four corner weeks, coldest (9--15 January 2021), darkest (28 November--4 December 2020), warmest (17--23 October 2020), and sunniest (13--19 February 2021).}
    \label{fig:extreme-violation-weeks}
\end{figure}

The four extreme weeks separate the controllers by operating condition and heating demand. 
In the cold and dark weeks, where heating demand is high, the data-driven controllers save a substantial amount of energy as compared to the hysteresis controller (Figure~\ref{fig:extreme-energy-weeks}), using around $190$--$196$\,kWh against $328$\,kWh in the coldest week. The margin is largest in exactly the conditions where anticipation matters most, since sustained high demand gives the predictive controllers the most opportunity to exploit the forecast.
In the warm and sunny weeks, the hysteresis controller reaches its switch-on threshold only occasionally, and then heats at full power, after which the mild conditions and the solar gains sustain the elevated temperature on their own. This keeps its energy use low, in the sunniest week $75$\,kWh against $81$--$109$\,kWh for the data-driven controllers, but leaves the room overheated. The considered data-based controllers never apply full power and stay close to the lower bound whenever they heat, which costs more energy but avoids the overheating, so the comfort violations of Select-DPC and GS-DPC are correspondingly lower (Figure~\ref{fig:extreme-violation-weeks}).
The warm and sunny weeks together account for only about $100$\,kWh, against roughly $2600$\,kWh over the full winter, so the differences in these low-demand weeks have limited bearing on seasonal energy, which is dominated by the cold and dark periods. Across the extreme weeks, the relations among the methods persist, with GS-DPC and Select-DPC giving the most consistent balance of energy and comfort and LPV-DPC the weakest on comfort.
   
\subsection{Comfort-Band Tightening}\label{sec:tightening}
Recall that in the nonlinear data-based predictive control formulations, we treat the comfort bound as a so-called soft constraint. This means that the optimization problem will naturally trade off constraint violation for energy saving, and hence, settle just below the lower bound of the comfort constraint. In this section, we show how this issue can be compensated for in a simple manner by tightening the comfort band, that is, feeding the optimization problem a conservative bound, and using the original bound for evaluation.
% New:
\rev{This shifts each predictive controller towards the comfort level of the hysteresis controller, so that the energy consumption can be compared at comparable comfort.}

Note that, while this section considers violations of the lower bound, violations of the upper bound are usually driven by solar gains and cannot be compensated by a heating controller. Hence, changing the upper bound in the cost function would have no effect.

\begin{figure*}[h!]
  \centering
  \includegraphics[width=0.8\textwidth]{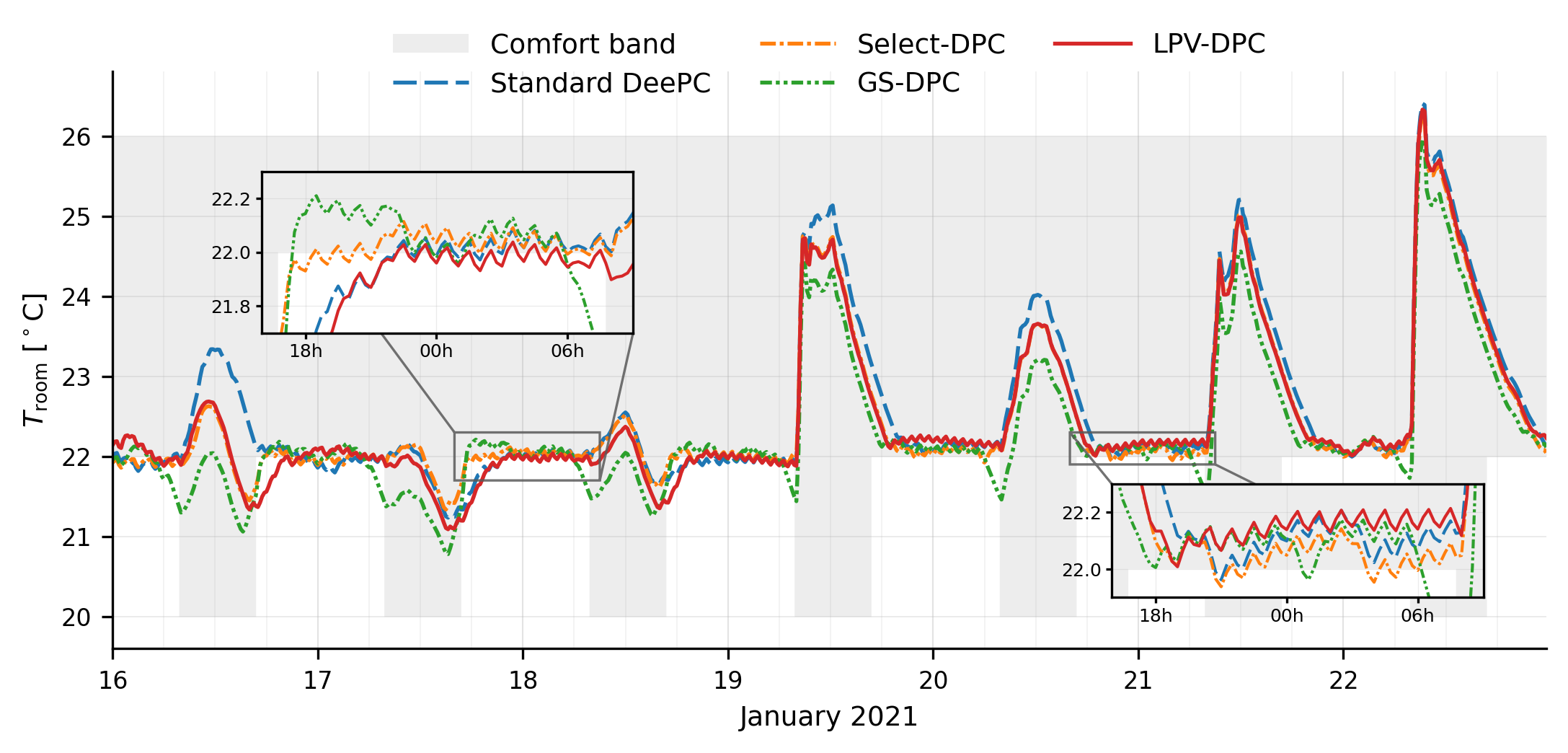}
    \caption{Room temperature of the four data-driven controllers with the tightened lower bound over the representative week (16--22 January 2021) in room R273, shown against the original comfort band.}
    \label{fig:tightened-behaviour}
\end{figure*}

For each controller and room, consider the undershoot error $e_k$ of the lower bound over the evaluation winter of Section~\ref{subsec:results},
\begin{equation}
e_k = \max\big(0,\ \underline{y}_k - y_k\big),
\end{equation}
Constraint tightening works by using the tightened lower bound
\begin{equation}
\underline{y}_k^{\delta} = \underline{y}_k + \delta,
\end{equation}
which replaces $\underline{y}_k$ in~\eqref{eq:band_slack}, where $\delta$ is the 90th percentile of the undershoot error $\{e_k\}_k$.
The values are listed in Table~\ref{tab:tightening}, with LPV-DPC requiring the largest tightening, consistent with its comfort violations in Section~\ref{subsec:results}.
\begin{table}[h!]
\centering
\footnotesize
\caption{Per-room lower-bound tightening $\delta$ in $^\circ$C, obtained as the 90th percentile of the undershoot $e_k$.}
\label{tab:tightening}
\begin{tabular*}{\columnwidth}{@{\extracolsep{\fill}}l rrr@{}}
\toprule
Controller & R272 & R273 & R274 \\
\midrule
DeePC & 0.08 & 0.21 & 0.00 \\
Select-DPC & 0.09 & 0.07 & 0.04 \\
GS-DPC & 0.07 & 0.18 & 0.02 \\
LPV-DPC & 0.31 & 0.33 & 0.11 \\
\bottomrule
\end{tabular*}
\end{table}
With the tightened bound, all four controllers reduce their seasonal comfort violation by between $21$ and $49\%$, at an energy cost of at most $2.6\%$ (Table~\ref{tab:winter-energy-comfort-q90}). A shift of the lower bound by at most a third of a degree therefore removes a substantial part of the violations.
\providecommand{\uu}[1]{{\footnotesize\,($\uparrow$#1\%)}}%
\providecommand{\dd}[1]{{\footnotesize\,($\downarrow$#1\%)}}%
\begin{table}[h!]
\centering
\footnotesize
\caption{Heating energy $E$ and comfort violation $V$ per controller over the evaluation winter with the tightened comfort reference, per room and summed over all rooms. The change column gives the difference in the total relative to the untightened setup of Table~\ref{tab:winter-energy-comfort}.}
\label{tab:winter-energy-comfort-q90}
\begin{tabular*}{\columnwidth}{@{\extracolsep{\fill}}l rrrrr@{}}
\toprule
Controller & R272 & R273 & R274 & Total & Change \\
\midrule
& \multicolumn{5}{c}{\textit{Energy} $E$ [kWh]} \\
\cmidrule(lr){2-6}
Hysteresis & 772 & 1577 & 523 & 2872 & --- \\
DeePC & 755 & 1279 & 691 & 2725 & $\uparrow$1.3\% \\
Select-DPC & 710 & 1366 & 490 & 2566 & $\uparrow$0.6\% \\
GS-DPC & 729 & 1373 & 502 & 2604 & $\uparrow$1.5\% \\
LPV-DPC & 716 & 1429 & 527 & 2672 & $\uparrow$2.6\% \\
\midrule
& \multicolumn{5}{c}{\textit{Comfort violation} $V$ [K\,h]} \\
\cmidrule(lr){2-6}
Hysteresis & 38 & 192 & 29 & 259 & --- \\
DeePC & 197 & 240 & 266 & 703 & $\downarrow$21.1\% \\
Select-DPC & 37 & 156 & 39 & 232 & $\downarrow$34.3\% \\
GS-DPC & 79 & 212 & 59 & 350 & $\downarrow$32.9\% \\
LPV-DPC & 166 & 374 & 158 & 699 & $\downarrow$49.4\% \\
\bottomrule
\end{tabular*}
\end{table}
All four controllers still consume less energy than the hysteresis controller, between $5.1$ and $10.7\%$, and Select-DPC now also violates the band less, $232$ against $259$\,K\,h. It is the only controller that outperforms the robust hysteresis controller on both metrics at once. Among the data-driven controllers, Select-DPC and GS-DPC remain ahead of standard DeePC in both energy and comfort. LPV-DPC gains the most from its tightening and now reaches the comfort level of standard DeePC while consuming less energy. Figure~\ref{fig:tightened-behaviour} shows the effect over the representative week, where the room temperature is now held at or slightly above the lower bound.
The tightening is an empirical quantile taken from a prior run of the same controller, so it presupposes that such results are available.
% New: 
\rev{Here, the quantile is taken from the evaluation winter itself, so the tightened results demonstrate how simply the comfort violations can be reduced rather than providing an independent performance estimate.}
It approximates a one-sided chance constraint, for which formal counterparts exist in stochastic and tube-based MPC~\citep{mesbah2016stochastic, MAYNE20142967}, and which have been carried over to DeePC, e.g., through distributionally robust and scenario-based formulations~\citep{coulson2021distributionally, zieglmeier2026scenario}. 
While such formulations offer more advanced constraint handling or stronger guarantees, integrating them into the DPC methods studied here is beyond the scope of this paper.

\subsection{Hyperparameters and Configuration}\label{sec:hp-tuning}
The hyperparameters are tuned separately for each controller and room by grid search during the week of 18--24 January 2021.
%New: 
\rev{This week lies within the evaluation period. Repeating the seasonal evaluation without it leaves all rankings in energy and comfort unchanged, as shown in Appendix~\ref{app:tuning-week}.}
They comprise the comfort and energy weights $Q$ and $R$, the regularization $\lambda_g$, and the slack weights $\lambda_{\sigma_y}$ and $\lambda_{\sigma_d}$, extended for LPV-DPC by the two scheduling-consistency slack weights $\lambda_{\sigma_y^\rho}$ and $\lambda_{\sigma_d^\rho}$, which are tuned separately and result in the same values as $\lambda_{\sigma_y}$ and $\lambda_{\sigma_d}$, respectively.
Candidates are ranked by the scalar score $E + 10\,V$, with the heating energy $E$ in kWh and the comfort violation $V$ in K\,h. A first coarse grid spanning several orders of magnitude locates the region of good performance, and each subsequent round refines the grid around the best cell of the previous one, down to a resolution at which the score changes only marginally.

The score alone is not sufficient to select the hyperparameters. In~\eqref{eq:deepc_dist}, one vector $g$ has to satisfy all blocks of the data equation simultaneously, so the predicted input $u = U_f\,g$ is produced by the same linear combination of columns that reproduces the forecast $\hat{d}$ in the future disturbance block. The slack $\sigma_d$, weighted by $\lambda_{\sigma_d}$, widens the set of admissible vectors $g$, but the input is still coupled to the disturbance. The hyperparameters are therefore additionally selected on their ability to withhold heating during the solar-gain phases of the day, representing sufficient decoupling, which is the behavior the energy savings depend on.

Several settings are shared by all four controllers. The comfort weight is $Q = 100$, and the horizons are $T_\mathrm{ini} = N = 8$, corresponding to two hours at the $15$-minute sampling time of the outer loop. The heating power is bounded by $0 \le u_k \le P_{\max}$ with $P_{\max} = \{2, 6, 2\}$\,kW for R272, R273 and R274, the actuation dead-band is set to $5\%$ of~$P_{\max}$, and the room temperature is constrained to $[10, 35]\,^\circ$C. The remaining values are used unchanged throughout the simulation study and are listed in Table~\ref{tab:hyperparameters}.
\begin{table}[t]
\centering
\footnotesize
\caption{Hyperparameters of the four data-driven controllers, with per-room entries listed as R272/R273/R274 and $M$ the number of columns entering the optimization.}
\label{tab:hyperparameters}
\begin{tabular*}{\columnwidth}{@{\extracolsep{\fill}}l ccccc@{}}
\toprule
Controller & $R$ & $\lambda_g$ & $\lambda_{\sigma_y}$ & $\lambda_{\sigma_d}$ & $M$ \\
\midrule
DeePC & 0.1 & 50/100/50 & $10^{3}$ & $10$ & 2628 \\
Select-DPC & 1/0.1/1 & 50/30/50 & $10^{4}$ & $10^{3}$ & 1000$^{\dagger}$ \\
GS-DPC & 0.1 & 5/10/5 & $10^{5}$ & $10^{3}$ & 377--1521$^{\ddagger}$ \\
LPV-DPC & 0.01 & 1/5/1 & $10^{3}$ & $10$ & 2628 \\
\bottomrule
\end{tabular*}\\[2pt]
{\footnotesize $^{\dagger}$Selected online at each step from a pool of $12\,483$ columns.\\
$^{\ddagger}$Partitioned offline into regions from a pool of $12\,483$ columns.}
\end{table}
The closed-loop simulations were carried out on the digital twin of Section~\ref{sec:digital-twin}, using EnergyPlus 9.3 and \textit{nestli} 0.1.0. The controllers were implemented in Python 3.10.11, with the underlying optimal control problems formulated in \texttt{CVXPY 1.7.5} and solved using \texttt{MOSEK 11.2.0}.

\subsection{Summary and Limitations}
\label{sec:results-summary}
No single controller dominates on both metrics in every operating condition, and several limitations of this study are discussed in the following. The data-driven controllers reduce seasonal heating energy compared to the hysteresis benchmark, most clearly in the high-demand cold and dark weeks that dominate winter consumption. Among them, the nonlinear methods improve on DeePC in energy throughout, and Select-DPC and GS-DPC do so in comfort as well, giving the best overall balance across scenarios. LPV-DPC, despite having the strongest theoretical foundation, does not translate this into comfort performance due to its sensitivity to noise, discussed in Section~\ref{sec:full-winter}. 

Across the predictive controllers, comfort violations follow from using the comfort band as a soft constraint, and the tightening method shown in Section~\ref{sec:tightening} mitigates them by simple means. As long as the band is treated as a soft constraint, however, comfort satisfaction cannot be guaranteed, only improved. The industry-standard hysteresis controller offers no such guarantees either, and violates the band as well due to the external disturbances acting on a real system, as reported above. 

Another limitation concerns the disturbances. In the simulation study, the controllers receive the exact future disturbance values over the prediction horizon rather than a forecast. This isolates the control formulations, the subject of this work, from forecast quality, which strongly affects closed-loop performance and would otherwise confound the comparison. In contrast, the real-world experiments presented in Section~\ref{sec:exp-results} do include these operational forecasts.

Practical deployment raises a different concern, since the hyperparameters detailed in Section~\ref{sec:hp-tuning} have to be manually fine-tuned. This means that the approach is not completely autonomous or plug-and-play. Automating this search is the missing piece towards deployment at scale. Differentiable tuning of DeePC hyperparameters~\citep{pmlr-v283-cummins25b} moves in this direction, although further development is needed before such methods reach the application stage. However, it should be noted that even with minimal tuning, all three nonlinear data-based predictive control methods outperform both linear methods and the baseline controller. 
%New:
\rev{Moreover, the tightening quantile is taken from the evaluation winter. Excluding the tuning week leaves all conclusions unchanged (Appendix~\ref{app:tuning-week}), but a held-out evaluation over a separate season would provide an independent estimate of the tightened performance.}

Furthermore, the data matrices are constructed from data that is collected offline and once per controller rather than updated during operation. The controllers can therefore not adapt to changes in the system, such as a renovation, a change in occupancy, or climate change. Hence, these controllers will need to be re-implemented with new data regularly. Adaptation of data-based predictive controllers using online measurements is an active area of research, and the present formulations are a natural starting point for such an extension.

Finally, Select-DPC occasionally encountered infeasibility during operation, $30$ times over the full evaluation winter across the three rooms and thus for $0.077\%$ of all solves, against $3$ for GS-DPC and none for LPV-DPC and standard DeePC. Because Select-DPC uses columns selected in terms of closeness to the current operating point, the selected data can become too similar to span the trajectory space, with the result that no feasible solution to the data equation exists. Whenever a solve failed, the heating input of the previous step was applied instead. These infeasibilities did not occur in succession, so this fallback was sufficient to maintain robust operation. 

This work establishes the foundation for applying nonlinear DPCs to heating systems, with a clear focus on applying nonlinear DPC to a real nonlinear building system and directly comparing the methods under identical conditions. Each of the limitations above builds on that basis and requires extensive research of its own, which is beyond the scope of this work.

%% -------------------------------------------------------
%% SECTION 6: EXPERIMENTAL RESULTS
%% -------------------------------------------------------

\section{Experimental Results}\label{sec:exp-results}
Access to the NEST demonstrator was limited to a four-week window in March and April 2026, and heating experiments additionally require outdoor conditions under which the unit is heated. Since March was unusually mild, the only suitable period arose when temperatures dropped again for some days at the turn of the month. The first part of this period was used to set up and commission the controller on the unit, leaving the interval from \formatdate{31}{3}{2026} 12:00 to \formatdate{4}{4}{2026} 00:00 for the evaluation reported below.
Both restrictions, the facility's availability and the weather, are inherent to experiments on an occupied residential heating system and cannot be resolved by planning alone.
Due to these limitations, the experiments were limited to a single method, and GS-DPC was selected as the most promising and safest method of nonlinear DPC. The control logic is identical to that of the simulation study, so the outer and inner loops of Section~\ref{sec:control-setup} operate on the physical unit exactly as they do on the digital twin.

Throughout the experiment, the research unit is occupied, and the shades and windows remain under the control of the residents, which act on the room as additional disturbances. In the simulation study of Section~\ref{sec:sim-results}, these disturbances are not modeled and held constant: the shades were left open, and the windows closed over both the data-collection and the evaluation period. Since an open window changes the air exchange directly, data recorded under this condition does not reflect the behavior for which the controller is built.
Data are extracted for each room over the period from \formatdate{1}{11}{2025} to \formatdate{31}{1}{2026}, and the intervals in which a window is open are removed. The remaining segments are combined into a mosaic-Hankel matrix, cf. Section~\ref{sec:data-generation}, so that no column contains a removed interval, and are then partitioned into the azimuth regions of Section~\ref{sec:GS-DPC}.
No excitation is designed for the real unit. The data are recorded under the hysteresis controller in normal operation, since only data already available from the running system can be used. The resulting input Hankel matrix has full row rank, suggesting sufficient excitation for the predictor. %, though the persistency-of-excitation guarantees apply only in the deterministic LTI setting and not exactly to a nonlinear unit.
The residents' use of shades and windows also determines which rooms admit a meaningful comparison. In Room~272 the shades remained fully closed throughout the experiments, and in Room~274 the window was opened frequently. In Room~273 both were used only occasionally, as visible in Figure~\ref{fig:exp_273}, so the following discussion focuses on this room.

\subsection{GS-DPC Configuration}\label{sec:exp-implementation}
\begin{figure*}
    \centering
    \includegraphics[width=0.8\linewidth]{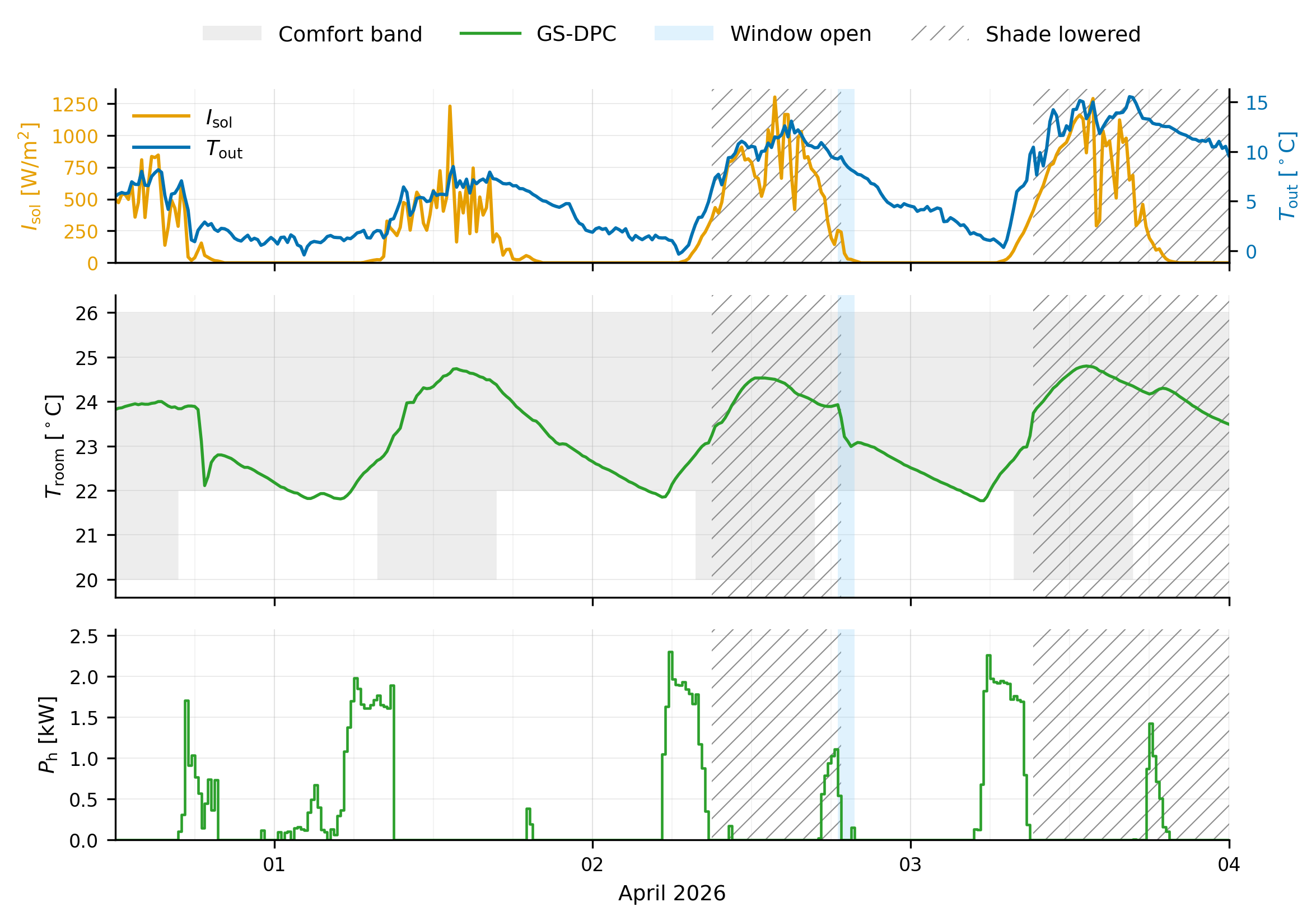}
    \caption{Closed-loop behavior of GS-DPC on the physical unit in room R273 over the experimental period (31 March--4 April 2026). Top: measured solar irradiance $I_\mathrm{sol}$ and outdoor temperature $T_\mathrm{out}$. Middle: room temperature $T_\mathrm{room}$ with the time-varying comfort band shaded. Bottom: heating power $P_\mathrm{h}$. Hatched areas mark lowered shades and the blue band an open window.}
    \label{fig:exp_273}
\end{figure*}

Tuning the hyperparameters on the physical unit is considerably more difficult than in simulation. The outer loop samples every $15$ minutes, and several iterations are needed before a change's effect becomes visible, so the search of Section~\ref{sec:hp-tuning} was infeasible within the experimental window. The degrees of freedom were therefore reduced before deployment.
The data-driven predictor of GS-DPC was first calibrated offline as a pure prediction model, using recorded heating input and disturbances from \formatdate{8}{2}{2026} to \formatdate{13}{2}{2026} to tune $\lambda_g$ and $\lambda_{\sigma_y}$ for prediction accuracy, without solving for an optimal input. This fixed the two weights relative to each other, leaving only their ratio to $Q$ and $R$ to tune online.
Two hyperparameter sets, \textit{v1} and \textit{v2}, were applied during the evaluation, listed in Table~\ref{tab:HP_exp}. On the first morning, the controller stopped heating abruptly and continued heating longer than the forecasted solar gain justified. Increasing the energy weight $R$ by an order of magnitude from \formatdate{1}{4}{2026} 19:00 onward corrected this, applying \textit{v2} from that point and producing the earlier and more gradual heating reduction visible in Figure~\ref{fig:exp_273}.
The resulting hyperparameters, reported in Section~\ref{sec:exp-results-discussion}, are not guaranteed to be optimal, but are sufficient to demonstrate that nonlinear DPC can control a real building effectively and robustly. Automated tuning, discussed in Section~\ref{sec:results-summary}, would be needed for large-scale deployment across many units.
The remaining settings follow the simulation study: a maximum heating power of $P_{\max} = \{2, 6, 2\}$\,kW for R272, R273 and R274, and room temperature constrained to $[10, 35]\,^\circ$C. The actuation dead-band differs from the simulation study and is set to $2\%$ of $P_{\max}$.
The controller was implemented in \texttt{Python 3.14.3}, with the optimal control problem formulated in \texttt{CVXPY 1.8.2} and solved using \texttt{MOSEK 11.1.10}, and communicated with the unit over \texttt{MQTT 3.1.1} using \texttt{paho-mqtt 2.1.0}.
\begin{table}[t]
\centering
\footnotesize
\caption{Hyperparameters of GS-DPC during the experiment, identical for all three rooms. The second set (\textit{v2}) was applied from \formatdate{1}{4}{2026} 19:00 onwards.}
\label{tab:HP_exp}
\begin{tabular*}{\columnwidth}{@{\extracolsep{\fill}}l ccccccc@{}}
\toprule
Set & $Q$ & $R$ & $\lambda_g$ & $\lambda_{\sigma_y}$ & $\lambda_{\sigma_d}$ & $T_\mathrm{ini}$ & $N$ \\
\midrule
\textit{v1} & $1000$ & $10$ & $10^{4}$ & $10^{5}$ & $10^{3}$ & $8$ & $8$ \\
\textit{v2} & $1000$ & $100$ & $10^{4}$ & $10^{5}$ & $10^{3}$ & $8$ & $8$ \\
\bottomrule
\end{tabular*}
\end{table}

The future disturbances are obtained through an interface provided by the NEST research environment, which supplies the operational weather forecast of the Swiss Federal Office of Meteorology and Climatology (MeteoSwiss). The forecast is updated four times a day and covers the following $72$ hours.
Another effect concerns the forecast itself. The past disturbances entering the optimization are measured values, whereas the future disturbances are forecasts, and the two are not always consistent at the boundary between them. Outdoor temperature evolves smoothly, so the recorded trajectories in the data matrices contain no abrupt changes in this channel. An inaccurate forecast, however, can differ substantially from the last measured value, producing a step at the past-to-future boundary that no linear combination of the recorded columns can reproduce.

The forecast is therefore blended with the last measurement. Over the prediction horizon, the outdoor-temperature entry is taken as the combination
\begin{equation}
T_{\mathrm{out},k} = \Big(1 - \frac{k}{N}\Big)\,\overline{T}_\mathrm{out} + \frac{k}{N}\,\hat{T}_{\mathrm{out},k}, \qquad k \in \mathbb{N}_{[0,\,N-1]},
\end{equation}
where $\overline{T}_\mathrm{out}$ is the outdoor temperature measured over the last control interval and $\hat{T}_{\mathrm{out},k}$ is the forecast. At the first step, the measurement carries full weight, so the trajectory is continuous across the boundary by construction, and the forecast gains influence as the prediction reaches further ahead, where it is more informative than persistence. The blending is applied to the outdoor temperature only, since solar radiation varies abruptly in the recorded data and therefore needs no smoothing.

\subsection{Results}\label{sec:exp-results-discussion}

The conditions during the experiment differ clearly from those of the simulation study. Outdoor temperatures reach up to $15\,^\circ$C and the solar irradiance exceeds $1200$\,W/m$^2$ on the clear days, so the heat entering the room through the windows covers a large part of the demand. The room temperature consequently remains in the upper half of the comfort band throughout, between roughly $22\,^\circ$C and $25\,^\circ$C, rather than tracking close to the lower bound as under the colder conditions of the simulation study.
The heating is concentrated in the night and early morning hours. The room cools once the solar gains cease, and GS-DPC applies heat as the temperature approaches the lower bound, which the occupancy schedule raises to $22\,^\circ$C at night. As soon as the disturbances warm the room again on the following morning, the controller stops heating and lets the solar gains provide enough heating throughout the day. This is the same behavior observed in the simulation study, here under conditions in which the passive gains cover a far larger share of the demand.

Each evening the controller applies a small amount of heat in azimuth regions $r=17$ to $r=19$, covering sunset, although the room temperature is comfortably inside the band and no heating is required.
A plausible explanation is the coupling between input and disturbance discussed in Section~\ref{sec:hp-tuning}. The hyperparameters are shared across all regions, even though GS-DPC holds a separate data set for each region. A hyperparameter configuration that decouples input and disturbance in most regions, need not do so in all of them. Confirming this would require per-region tuning, which multiplies the tuning effort by the number of regions and strengthens the case for the automated search discussed in Section~\ref{sec:results-summary}.

The interventions of the residents are visible in the same figure. The shades are lowered on both clear days, and the brief window opening on \formatdate{2}{4}{2026} produces an immediate drop in room temperature. The controller handles both robustly without leaving the comfort band.

\subsection{Summary and Limitations}\label{sec:exp-summary}
%The deployment demonstrates that GS-DPC operates robustly on a real, occupied residential unit under disturbances, matching the control performance observed in the simulation study of Section~\ref{sec:sim-results}.
% Replaced by: 
\rev{The deployment demonstrates that GS-DPC operates robustly on a real, occupied residential unit under disturbances, with the same qualitative behavior as observed in the simulation study of Section~\ref{sec:sim-results}. Since the experiment covers a short period under mild conditions and includes no benchmark controller, it does not quantify energy savings, which are assessed in the simulation study.}
The experiments were conducted \emph{before} the simulation study. While this would not be the favorable methodological order when a calibrated twin is available (since the twin would otherwise inform the experimental design), it does reflect the reality of performing experimental research. The described order is thus a result of the availability of the research unit and weather restrictions rather than a methodological preference.

In a real-world setting, obtaining well-tuned hyperparameters is a main limitation for the reasons given in Section~\ref{sec:exp-implementation}. The resulting hyperparameters used in the experiment, while not necessarily optimal, were nonetheless sufficient for GS-DPC to control the unit effectively and robustly, as reported in Section~\ref{sec:exp-results-discussion}. Incorporating the operational forecast (the blending introduced in Section~\ref{sec:exp-implementation}) was achieved robustly by a simple, minor modification.
A further limitation concerns the data itself. Unlike the simulation study, no open-loop experiments for data collection could be implemented in the occupied apartment. The data matrices instead had to be constructed from historical operation under the hysteresis controller. This still yielded effective control performance. Adapting to long-term changes in the system, as discussed in Section~\ref{sec:results-summary}, would require updating the data set online with data collected under the deployed DPCs. This would result in an adaptive DPC form. Adaptive DPC is an active area of research, and its theoretical foundations are not yet developed to the point of practical application.

%% -------------------------------------------------------
%% SECTION 7: CONCLUSION
%% -------------------------------------------------------
\section{Conclusion}\label{sec:conclusion}
Residential heating is nonlinear and heterogeneous, which limits MPC to settings in which the effort of building and maintaining a model per unit can be justified. DeePC removes this step, but its foundation restricts it to deterministic LTI systems. This work brought three nonlinear methods of the method, Select-DPC, GS-DPC, and LPV-DPC, together with standard DeePC and a hysteresis benchmark, onto a residential heating problem and compared them over a full heating season on a calibrated digital twin of an occupied research unit.
% All data-driven controllers consume less energy than the hysteresis benchmark, with Select-DPC saving roughly $11\%$ over the season, and the margin...
% Replaced by
All data-driven controllers consume less energy than the hysteresis benchmark, \rev{at the cost of higher comfort-band violations,} and the margin is largest in the cold and dark weeks that dominate winter consumption. The three nonlinear methods consume less energy than standard DeePC throughout, and Select-DPC and GS-DPC also violate the comfort band less, giving the best balance of the two metrics. LPV-DPC does not translate its theoretical foundation into comfort performance under these noisy conditions. Because the comfort band enters the cost rather than the constraints, all predictive controllers trade comfort against energy. Tightening the lower comfort bound reduces the seasonal violations by between $21\%$ and $49\%$ at an energy cost of at most $2.6\%$.
% New: 
\rev{With the tightened bound, Select-DPC saves $10.7\%$ of the heating energy while violating the comfort band less than the hysteresis benchmark.}
GS-DPC was deployed on the physical unit and operated on the operational weather forecast. It held the room temperature within the comfort band throughout. To the best of our knowledge, this is the first deployment of a nonlinear DPC method on a real heating system, 
% and it shows that these methods control residential heating reliably without a building model.
% New: 
and it shows that \rev{GS-DPC controls} residential heating reliably without a building model.

Three further research directions follow from this. The hyperparameters still require manual tuning, which is the remaining obstacle to deployment across a building stock at large scale, and GS-DPC raises the further question of whether each scheduling region needs its own configuration. The comfort band is enforced only through the cost, so replacing the empirical tightening by a chance-constrained, robust, or scenario-based formulation would allow comfort to be treated as a constraint. The data matrices are built once offline, and updating them during operation would let the controllers follow changes in the system.

% TODO: ~0.5 page + references
\printcredits
\subsection*{Declaration of generative AI and AI-assisted technologies in the manuscript preparation process}
During the preparation of this work the author(s) used Claude in order to assist with writing. After using this tool/service, the author(s) reviewed and edited the content as needed and take(s) full responsibility for the content of the published article.

\appendix
\section{\rev{Supplementary Evaluation}}\label{app:supplementary}

\subsection{\rev{Seasonal Evaluation without the Tuning Week}}\label{app:tuning-week}
\rev{The hyperparameters of Section~\ref{sec:hp-tuning} are tuned on the week of 18--24 January 2021, which lies within the evaluation winter. Since the predictor is formed directly from the data of winter 2019--2020, no model is fitted to the evaluation data, and only the weights of the cost function are chosen on this week. To assess the influence of this overlap, Tables~\ref{tab:app-winter} and~\ref{tab:app-winter-q90} repeat the seasonal evaluation of Tables~\ref{tab:winter-energy-comfort} and~\ref{tab:winter-energy-comfort-q90} without the tuning week, over the remaining $128$ days. The rankings in energy and comfort are identical to those over the full winter, for both the original and the tightened comfort band. With the tightened band, Select-DPC consumes $10.7\%$ less energy than the hysteresis controller while violating the comfort band less ($225$ versus $239$\,K\,h), and the tightening reduces the violations by $21$--$48\%$ at an energy increase of at most $2.6\%$.}

\begin{table}[]
%\color{blue}
\centering
\footnotesize
\setlength{\tabcolsep}{3pt}
\caption{Heating energy and comfort-band violation per controller over the evaluation winter without the tuning week (17 October 2020--28 February 2021, excluding 18--24 January 2021), per controlled room and summed over all three rooms.}
\label{tab:app-winter}
\begin{tabular}{l rrrr rrrr}
\toprule
& \multicolumn{4}{c}{$E$ [kWh]} & \multicolumn{4}{c}{$V$ [K\,h]} \\
\cmidrule(lr){2-5}\cmidrule(lr){6-9}
Controller & R272 & R273 & R274 & Total & R272 & R273 & R274 & Total \\
\midrule
Hysteresis & 735 & 1497 & 495 & 2727 & 36 & 176 & 27 & 239 \\
DeePC & 719 & 1168 & 657 & 2543 & 234 & 383 & 256 & 873 \\
Select-DPC & 665 & 1291 & 464 & 2420 & 99 & 185 & 56 & 340 \\
GS-DPC & 698 & 1256 & 481 & 2436 & 115 & 327 & 62 & 503 \\
LPV-DPC & 666 & 1296 & 504 & 2466 & 463 & 657 & 216 & 1335 \\
\bottomrule
\end{tabular}
\end{table}

\providecommand{\uu}[1]{{\footnotesize\,($\uparrow$#1\%)}}%
\providecommand{\dd}[1]{{\footnotesize\,($\downarrow$#1\%)}}%
\begin{table}[h!]
%\color{blue}
\centering
\footnotesize
\caption{Heating energy $E$ and comfort violation $V$ per controller with the tightened comfort reference over the evaluation winter without the tuning week, per room and summed over all rooms. The change column gives the difference in the total relative to the untightened setup of Table~\ref{tab:app-winter}.}
\label{tab:app-winter-q90}
\begin{tabular*}{\columnwidth}{@{\extracolsep{\fill}}l rrrrr@{}}
\toprule
Controller & R272 & R273 & R274 & Total & Change \\
\midrule
& \multicolumn{5}{c}{\textit{Energy} $E$ [kWh]} \\
\cmidrule(lr){2-6}
Hysteresis & 735 & 1497 & 495 & 2727 & --- \\
DeePC & 714 & 1212 & 653 & 2579 & $\uparrow$1.4\% \\
Select-DPC & 672 & 1298 & 463 & 2434 & $\uparrow$0.6\% \\
GS-DPC & 693 & 1306 & 473 & 2473 & $\uparrow$1.5\% \\
LPV-DPC & 678 & 1354 & 497 & 2530 & $\uparrow$2.6\% \\
\midrule
& \multicolumn{5}{c}{\textit{Comfort violation} $V$ [K\,h]} \\
\cmidrule(lr){2-6}
Hysteresis & 36 & 176 & 27 & 239 & --- \\
DeePC & 195 & 233 & 263 & 691 & $\downarrow$20.8\% \\
Select-DPC & 37 & 150 & 38 & 225 & $\downarrow$33.7\% \\
GS-DPC & 78 & 206 & 58 & 342 & $\downarrow$32.1\% \\
LPV-DPC & 166 & 367 & 157 & 690 & $\downarrow$48.3\% \\
\bottomrule
\end{tabular*}
\end{table}

\subsection{\rev{Commanded and Delivered Heating Power}}\label{app:power}
\rev{The inner loop of Section~\ref{sec:control-setup} realizes the commanded heating power $u_k$ by pulse-width modulation of the valve. Since NEST operates on a one-minute resolution, the valve openings can be realized only in full minutes, so that the delivered power $P_{\mathrm{h},k}$ exceeds the commanded power. Table~\ref{tab:power-discrepancy} lists the mean surplus $\Delta\bar{P}$ of the delivered over the commanded power, averaged over all active control intervals with $u_k > 0$, for each predictive controller and room. The surplus amounts to about $60$\,W in R272 and R274 and about $200$\,W in R273, corresponding to $3.2$--$3.4\%$ of the respective maximum power $P_{\max}$. It is one-sided, with the delivered power falling below the commanded power in only about $0.3\%$ of the active intervals, and nearly identical for all four predictive controllers. The hysteresis controller opens the valve over entire control intervals and is not affected by this resolution. All energies reported in this work are computed from the delivered power and therefore include this surplus.}

\begin{table}[h!]
%\color{blue}
\centering
\footnotesize
\caption{Mean surplus $\Delta\bar{P}$ of the delivered over the commanded heating power per predictive controller and room, averaged over all active control intervals of the evaluation winter, and the surplus relative to the maximum power $P_{\max}$, pooled over all rooms.}
\label{tab:power-discrepancy}
\begin{tabular*}{\columnwidth}{@{\extracolsep{\fill}}l rrrr@{}}
\toprule
& \multicolumn{3}{c}{$\Delta\bar{P}$ [W]} & \\
\cmidrule(lr){2-4}
Controller & R272 & R273 & R274 & $\Delta\bar{P}/P_{\max}$ [\%] \\
\midrule
& \multicolumn{4}{c}{\textit{Original comfort band}} \\
\cmidrule(lr){2-5}
DeePC & 64 & 206 & 64 & 3.3 \\
Select-DPC & 64 & 192 & 65 & 3.2 \\
GS-DPC & 64 & 203 & 64 & 3.2 \\
LPV-DPC & 67 & 212 & 68 & 3.4 \\
\midrule
& \multicolumn{4}{c}{\textit{Tightened comfort band}} \\
\cmidrule(lr){2-5}
DeePC & 63 & 202 & 62 & 3.2 \\
Select-DPC & 63 & 194 & 64 & 3.2 \\
GS-DPC & 64 & 196 & 64 & 3.2 \\
LPV-DPC & 66 & 207 & 67 & 3.4 \\
\bottomrule
\end{tabular*}
\end{table}
\FloatBarrier
\bibliographystyle{cas-model2-names}

\bibliography{MyBib}

@techreport{iea2023buildings,
  author      = {{International Energy Agency}},
  title       = {Energy Efficiency 2023},
  institution = {IEA},
  year        = {2023},
  address     = {Paris},
  note        = {\url{https://www.iea.org/reports/energy-efficiency-2023}}
}

@article{oldewurtel2012use,
  title={Use of model predictive control and weather forecasts for energy efficient building climate control},
  author={Oldewurtel, Frauke and Parisio, Alessandra and Jones, Colin N and Gyalistras, Dimitrios and Gwerder, Markus and Stauch, Vanessa and Lehmann, Beat and Morari, Manfred},
  journal={Energy and Buildings},
  volume={45},
  number={},
  pages={15--27},
  year={2012},
  publisher={Elsevier}
}

@article{drgovna2020all,
  title={All you need to know about model predictive control for buildings},
  author={Drgo{\v{n}}a, J{\'a}n and Arroyo, Javier and Figueroa, Iago Cupeiro and Blum, David and Arendt, Krzysztof and Kim, Donghun and Oll{\'e}, Enric Perarnau and Oravec, Juraj and Wetter, Michael and Vrabie, Draguna L and others},
  journal={Annual Reviews in Control},
  volume={50},
  number={},
  pages={190--232},
  year={2020},
  publisher={Elsevier}
}

@inproceedings{coulson2019data,
  title={Data-enabled predictive control: In the shallows of the {{DeePC}}},
  author={Coulson, Jeremy and Lygeros, John and D{\"o}rfler, Florian},
  booktitle={2019 18th European Control Conference (ECC)},
  pages={307--312},
  year={2019},
  organization={IEEE} 
}

@article{willems2005note,
  title={A note on persistency of excitation},
  author={Willems, Jan C and Rapisarda, Paolo and Markovsky, Ivan and De Moor, Bart LM},
  journal={Systems \& Control Letters},
  volume={54},
  number={4},
  pages={325--329},
  year={2005},
  publisher={Elsevier}
}

@article{dorfler2022bridging,
  title={Bridging direct and indirect data-driven control formulations via regularizations and relaxations},
  author={D{\"o}rfler, Florian and Coulson, Jeremy and Markovsky, Ivan},
  journal={IEEE Transactions on Automatic Control},
  volume={68},
  number={2},
  pages={883--897},
  year={2022},
  publisher={IEEE} 
}

@incollection{verhoek2024encyclo,
	author = {Roland T{\'o}th and Chris Verhoek},
	booktitle = {Encyclopedia of Systems and Control Engineering},
	volume = {1},
    number = {},
	pages = {405-418},
	publisher = {Elsevier},
	title = {Modeling and Control of {LPV} systems},
	year = {2026}
}

@article{verhoek2021data,
  title={Data-driven predictive control for linear parameter-varying systems},
  author={Verhoek, Chris and Abbas, Hossam S and T{\'o}th, Roland and Haesaert, Sofie},
  journal={IFAC-PapersOnLine},
  volume={54},
  number={8},
  pages={101--108},
  year={2021},
  publisher={Elsevier} 
}

@inproceedings{verhoek2021fundamental,
  title={Fundamental lemma for data-driven analysis of linear parameter-varying systems},
  author={Verhoek, Chris and T{\'o}th, Roland and Haesaert, Sofie and Koch, Anne},
  booktitle={2021 60th IEEE Conference on Decision and Control (CDC)},
  pages={5040--5046},
  year={2021},
  organization={IEEE} 
}

@article{verhoek2025linear,
  title={A linear parameter-varying approach to data predictive control},
  author={Verhoek, Chris and Berberich, Julian and Haesaert, Sofie and T{\'o}th, Roland and Abbas, Hossam S},
  journal={IEEE Transactions on Automatic Control},
  year={2026},
  volume={71},
  number={4},
  pages={2512-2527},
  publisher={IEEE} 
}

@article{verhoek2025behavioural,
  author={Verhoek, Chris and Markovsky, Ivan and Haesaert, Sofie and Tóth, Roland},
  journal={IEEE Transactions on Automatic Control}, 
  title={A Behavioral Approach for {LPV} Data-Driven Representations}, 
  year={2026},
  volume={71},
  number={3},
  pages={1616-1629}
  }

@article{naf2025choose,
  title={Choose wisely: Data-driven predictive control for nonlinear systems using online data selection},
  author={N{\"a}f, Joshua and Moffat, Keith and Eising, Jaap and D{\"o}rfler, Florian},
  journal={arXiv preprint},
  year={2026} 
}

@article{zieglmeier2025gain,
  title={Gain-Scheduling Data-Enabled Predictive Control for Nonlinear Systems with Linearized Operating Regions},
  author={Zieglmeier, Sebastian and Hudoba de Badyn, Mathias and Warakagoda, Narada D and Krogstad, Thomas R and Engelstad, Paal},
  journal={arXiv preprint},
  year={2026} 
}

@article{guerrero2025gain,
  title={Gain-Scheduled Data-Enabled Predictive Control: A {{DeePC}} Approach for Nonlinear Systems},
  author={Guerrero, Margarita A and Lakshminarayanan, Braghadeesh and Rojas, Cristian R},
  journal={IEEE Control Systems Letters},
  volume={9},
  number={},
  pages={3041--3046},
  year={2025},
  publisher={IEEE} 
}

@article{markovsky2022identifiability,
  title={Identifiability in the behavioral setting},
  author={Markovsky, Ivan and D{\"o}rfler, Florian},
  journal={IEEE Transactions on Automatic Control},
  volume={68},
  number={3},
  pages={1667--1677},
  year={2022},
  publisher={IEEE} 
}

@article{markovsky2021behavioral,
  title={Behavioral systems theory in data-driven analysis, signal processing, and control},
  author={Markovsky, Ivan and D{\"o}rfler, Florian},
  journal={Annual Reviews in Control},
  volume={52},
  number={},
  pages={42--64},
  year={2021},
  publisher={Elsevier} 
}

@article{verheijen2023handbook,
  title={Handbook of linear data-driven predictive control: Theory, implementation and design},
  author={Verheijen, PCN and Breschi, Valentina and Lazar, Mircea},
  journal={Annual Reviews in Control},
  volume={56},
  number={},
  pages={100914},
  year={2023},
  publisher={Elsevier} 
}

@article{yin2024data,
  title={Data-driven predictive control for demand side management: Theoretical and experimental results},
  author={Yin, Mingzhou and Cai, Hanmin and Gattiglio, Andrea and Khayatian, Fazel and Smith, Roy S and Heer, Philipp},
  journal={Applied Energy},
  volume={353},
  number={},
  pages={122101},
  year={2024},
  publisher={Elsevier} 
}

@article{beerwerth2025morecontextualsamplingnonlinear,
  title={Less is More: Contextual Sampling for Nonlinear Data-Driven Predictive Control},
  author={Beerwerth, Julius and Alrifaee, Bassam},
  journal={arXiv preprint},
  year={2025} 
}

@article{berberich2020data,
  title={Data-driven model predictive control with stability and robustness guarantees},
  author={Berberich, Julian and K{\"o}hler, Johannes and M{\"u}ller, Matthias A and Allg{\"o}wer, Frank},
  journal={IEEE Transactions on Automatic Control},
  volume={66},
  number={4},
  pages={1702--1717},
  year={2020},
  publisher={IEEE}
}

@article{berberich2021data,
  title={Data-driven model predictive control: closed-loop guarantees and experimental results},
  author={Berberich, Julian and K{\"o}hler, Johannes and M{\"u}ller, Matthias A and Allg{\"o}wer, Frank},
  journal={at-Automatisierungstechnik},
  volume={69},
  number={7},
  pages={608--618},
  year={2021},
  publisher={De Gruyter Oldenbourg} 
}

@article{richner2017nest,
  title={{{NEST}} -- a platform for the acceleration of innovation in buildings},
  author={Richner, Peter and Heer, Ph and Largo, Reto and Marchesi, Enrico and Zimmermann, Mark},
  journal={Informes de la Construcci{\'o}n},
  volume={69},
  number={548},
  pages={e222},
  year={2017},
  publisher={Consejo Superior de Investigaciones Cient{\'\i}ficas}
}

@article{bojarski2023nestli,
  title={nestli: Neighborhood Energy System Testing towards Large-scale Integration},
  author={Bojarski, Aaron and Khayatian, Fazel and Cai, Hanmin},
  journal={Zenodo},
  year={2023} 
}

@inproceedings{giacomelli2025insights,
  title={Insights into the explainability of {{Lasso}}-based {{DeePC}} for nonlinear systems},
  author={Giacomelli, Gianluca and Formentin, Simone and Lopez, Victor G and M{\"u}ller, Matthias A and Breschi, Valentina},
  booktitle={2025 IEEE 64th Conference on Decision and Control (CDC)},
  pages={1036--1041},
  year={2025},
  organization={IEEE} 
}

@article{rugh2000research,
  title={Research on gain scheduling},
  author={Rugh, Wilson J and Shamma, Jeff S},
  journal={Automatica},
  volume={36},
  number={10},
  pages={1401--1425},
  year={2000},
  publisher={Elsevier} 
}

@article{van2020willems,
  title={Willems’ fundamental lemma for state-space systems and its extension to multiple datasets},
  author={Van Waarde, Henk J and De Persis, Claudio and Camlibel, M Kanat and Tesi, Pietro},
  journal={IEEE Control Systems Letters},
  volume={4},
  number={3},
  pages={602--607},
  year={2020},
  publisher={IEEE}  
}

@article{crawley2001energyplus,
  title={{{EnergyPlus}}: creating a new-generation building energy simulation program},
  author={Crawley, Drury B and Lawrie, Linda K and Winkelmann, Frederick C and Buhl, Walter F and Huang, Y Joe and Pedersen, Curtis O and Strand, Richard K and Liesen, Richard J and Fisher, Daniel E and Witte, Michael J and others},
  journal={Energy and Buildings},
  volume={33},
  number={4},
  pages={319--331},
  year={2001},
  publisher={Elsevier}
}

@inproceedings{heisel2019resource,
  title={Resource-respectful construction--the case of the {{Urban Mining and Recycling}} unit ({{UMAR}})},
  author={Heisel, Felix and Hebel, Dirk E and Sobek, Werner},
  booktitle={IOP Conference Series: Earth and Environmental Science},
  volume={225},
  number={1},
  pages={012049},
  year={2019},
  organization={IOP Publishing} 
}

@inproceedings{sturzenegger2014brcm,
  title={{{BRCM}} {{Matlab}} Toolbox: Model generation for model predictive building control},
  author={Sturzenegger, David and Gyalistras, Dimitrios and Semeraro, Vito and Morari, Manfred and Smith, Roy S},
  booktitle={2014 American Control Conference},
  pages={1063--1069},
  year={2014},
  organization={IEEE} 
}

@inproceedings{khayatian2022benchmarking,
  title={Benchmarking {{HVAC}} controller performance with a digital twin},
  author={Khayatian, Fazel and Cai, Hanmin and Bojarski, Aaron and Heer, Philipp and Bollinger, Andrew},
  booktitle={Applied Energy Symposium},
  volume={31},
  year={2022} 
}

@article{mesbah2016stochastic,
  title={Stochastic model predictive control: An overview and perspectives for future research},
  author={Mesbah, Ali},
  journal={IEEE Control Systems Magazine},
  volume={36},
  number={6},
  pages={30--44},
  year={2016},
  publisher={IEEE}
}

@article{MAYNE20142967,
    title = {Model predictive control: Recent developments and future promise},
    journal = {Automatica},
    volume = {50},
    number = {12},
    pages = {2967-2986},
    year = {2014},
    issn = {0005-1098},
    author = {David Q. Mayne}
}

@article{coulson2021distributionally,
  title={Distributionally robust chance constrained data-enabled predictive control},
  author={Coulson, Jeremy and Lygeros, John and D{\"o}rfler, Florian},
  journal={IEEE Transactions on Automatic Control},
  volume={67},
  number={7},
  pages={3289--3304},
  year={2021},
  publisher={IEEE} 
}

@article{zieglmeier2026scenario,
  title={Scenario-based Data-Enabled Predictive Control: Robustification via the Scenario Approach},
  author={Zieglmeier, Sebastian and Recke, Nikolas and Hudoba de Badyn, Mathias},
  journal={arXiv preprint},
  year={2026} 
}

@article{elokda2021data,
  title={Data-enabled predictive control for quadcopters},
  author={Elokda, Ezzat and Coulson, Jeremy and Beuchat, Paul N and Lygeros, John and D{\"o}rfler, Florian},
  journal={International Journal of Robust and Nonlinear Control},
  volume={31},
  number={18},
  pages={8916--8936},
  year={2021},
  publisher={Wiley Online Library} 
}

@article{zieglmeier2025auv,
  title={Data-Enabled Predictive Control with Predictive Adaptive Line-of-Sight Guidance for 3-D Path Following of Autonomous Underwater Vehicles},
  author={Zieglmeier, Sebastian and Hudoba de Badyn, Mathias and Warakagoda, Narada D and Krogstad, Thomas R and Engelstad, Paal},
  journal={arXiv preprint},
  year={2026}
}

@article{stoffel2023evaluation,
  title={Evaluation of advanced control strategies for building energy systems},
  author={Stoffel, Phillip and Maier, Laura and K{\"u}mpel, Alexander and Schreiber, Thomas and M{\"u}ller, Dirk},
  journal={Energy and Buildings},
  volume={280},
  number={},
  pages={112709},
  year={2023},
  publisher={Elsevier} 
}

@phdthesis{Verhoek_Thesis,
  author = {Chris Verhoek},
  title = {{Data-Driven Analysis and Control of Nonlinear Systems with Stability and Performance Guarantees: A Linear Parameter-Varying Approach}},
  school = {Eindhoven University of Technology, The Netherlands},
  year = {2025}
}

@InProceedings{pmlr-v283-cummins25b,
  title = 	 {{{DeePC}}-{{Hunt}}: Data-enabled Predictive Control Hyperparameter Tuning via Differentiable Optimization},
  author =       {Cummins, Michael and Padoan, Alberto and Moffat, Keith and Dorfler, Florian and Lygeros, John},
  booktitle = 	 {Proceedings of the 7th Annual Learning for Dynamics \& Control Conference},
  pages = 	 {673--685},
  year = 	 {2025},
  volume = 	 {283},
  series = 	 {Proceedings of Machine Learning Research},
  month = 	 {04--06 Jun},
  publisher =    {PMLR}
}

@article{cai2024experimental,
  title={Experimental implementation of an emission-aware prosumer with online flexibility quantification and provision},
  author={Cai, Hanmin and Heer, Philipp},
  journal={Sustainable Cities and Society},
  volume={111},
  number={},
  pages={105531},
  year={2024},
  publisher={Elsevier} 
}

@ARTICLE{10089206,
  author={Lian, Yingzhao and Shi, Jicheng and Koch, Manuel and Jones, Colin Neil},
  journal={IEEE Transactions on Control Systems Technology}, 
  title={Adaptive Robust Data-Driven Building Control via Bilevel Reformulation: An Experimental Result}, 
  year={2023},
  volume={31},
  number={6},
  pages={2420-2436}
}

@INPROCEEDINGS{9992445,
  author={Di Natale, Loris and Lian, Yingzhao and Maddalena, Emilio T. and Shi, Jicheng and Jones, Colin N.},
  booktitle={2022 IEEE 61st Conference on Decision and Control (CDC)}, 
  title={Lessons Learned from Data-Driven Building Control Experiments: Contrasting {{Gaussian}} Process-based {{MPC}}, Bilevel {{DeePC}}, and Deep Reinforcement Learning}, 
  year={2022},
  volume={},
  number={},
  pages={1111-1117}
}

@article{BUNNING2020109792,
title = {Experimental demonstration of data predictive control for energy optimization and thermal comfort in buildings},
journal = {Energy and Buildings},
volume = {211},
number={},
pages = {109792},
year = {2020},
issn = {0378-7788},
author = {Felix Bünning and Benjamin Huber and Philipp Heer and Ahmed Aboudonia and John Lygeros}
}

@inproceedings{di2021deep,
  title={Deep Reinforcement Learning for room temperature control: a black-box pipeline from data to policies},
  author={Di Natale, Loris and Svetozarevic, Bratislav and Heer, Philipp and Jones, CN},
  booktitle={Journal of Physics: Conference Series},
  volume={2042},
  number={1},
  pages={012004},
  year={2021},
  organization={IOP Publishing} 
}

@article{berberich2022linear,
  title={Linear tracking {{MPC}} for nonlinear systems—{{Part II}}: The data-driven case},
  author={Berberich, Julian and K{\"o}hler, Johannes and M{\"u}ller, Matthias A and Allg{\"o}wer, Frank},
  journal={IEEE Transactions on Automatic Control},
  volume={67},
  number={9},
  pages={4406--4421},
  year={2022},
  publisher={IEEE}  
}

@article{huang2023robust,
  title={Robust and kernelized data-enabled predictive control for nonlinear systems},
  author={Huang, Linbin and Lygeros, John and D{\"o}rfler, Florian},
  journal={IEEE Transactions on Control Systems Technology},
  volume={32},
  number={2},
  pages={611--624},
  year={2023},
  publisher={IEEE}  
}

@article{lian2021koopman,
  title={Koopman based data-driven predictive control},
  author={Lian, Yingzhao and Wang, Renzi and Jones, Colin N},
  journal={arXiv preprint},
  year={2021} 
}

@inproceedings{lian2021nonlinear,
  title={Nonlinear data-enabled prediction and control},
  author={Lian, Yingzhao and Jones, Colin N},
  booktitle={Learning for Dynamics and Control},
  pages={523--534},
  year={2021},
  organization={PMLR}
}

@inproceedings{lazar2024basis,
  title={Basis-functions nonlinear data-enabled predictive control: Consistent and computationally efficient formulations},
  author={Lazar, Mircea},
  booktitle={2024 European Control Conference (ECC)},
  pages={888--893},
  year={2024},
  organization={IEEE}  
}

@article{giacomelli2026beyond,
  title={Beyond Shrinkage: Foundations of Data-Driven Control for Piecewise Affine Systems},
  author={Giacomelli, Gianluca and Lopez, Victor G and Formentin, Simone and M{\"u}ller, Matthias A and Breschi, Valentina},
  journal={arXiv preprint},
  year={2026} 
}

@article{engeln2026data,
  title={Data-driven predictive control of nonlinear systems using weighted regularization},
  author={Engeln, Fritz A and Zieglmeier, Sebastian and Zag{\'o}rowska, Marta and van Wingerden, Jan-Willem},
  journal={arXiv preprint},
  year={2026} 
}

@article{svetozarevic2022data,
  title={Data-driven control of room temperature and bidirectional {{EV}} charging using deep reinforcement learning: Simulations and experiments},
  author={Svetozarevic, Bratislav and Baumann, Christian and Muntwiler, Simon and Di Natale, Loris and Zeilinger, Melanie N and Heer, Philipp},
  journal={Applied Energy},
  volume={307},
  number={},
  pages={118127},
  year={2022},
  publisher={Elsevier} 
}

@inproceedings{di2022near,
  title={Near-optimal deep reinforcement learning policies from data for zone temperature control},
  author={Di Natale, Loris and Svetozarevic, Bratislav and Heer, Philipp and Jones, Colin N},
  booktitle={2022 IEEE 17th International Conference on Control \& Automation (ICCA)},
  pages={698--703},
  year={2022},
  organization={IEEE} 
}

@inproceedings{bunning2021input,
  title={Input convex neural networks for building {{MPC}}},
  author={B{\"u}nning, Felix and Schalbetter, Adrian and Aboudonia, Ahmed and Hudoba de Badyn, Mathias and Heer, Philipp and Lygeros, John},
  booktitle={Learning for Dynamics and Control},
  pages={251--262},
  year={2021},
  organization={PMLR}
}

@inproceedings{zieglmeier2025semi,
  title={Semi-data-driven model predictive control: A physics-informed data-driven control approach},
  author={Zieglmeier, Sebastian and Hudoba de Badyn, Mathias and Warakagoda, Narada D and Krogstad, Thomas R and Engelstad, Paal},
  booktitle={2025 IEEE 64th Conference on Decision and Control (CDC)},
  pages={2110--2117},
  year={2025},
  organization={IEEE} 
}

@article{bunning2022physics,
  title={Physics-informed linear regression is competitive with two Machine Learning methods in residential building {{MPC}}},
  author={B{\"u}nning, Felix and Huber, Benjamin and Schalbetter, Adrian and Aboudonia, Ahmed and Hudoba de Badyn, Mathias and Heer, Philipp and Smith, Roy S and Lygeros, John},
  journal={Applied Energy},
  volume={310},
  number={},
  pages={118491},
  year={2022},
  publisher={Elsevier} 
}

@article{yang2020model,
  title={Model predictive control with adaptive machine-learning-based model for building energy efficiency and comfort optimization},
  author={Yang, Shiyu and Wan, Man Pun and Chen, Wanyu and Ng, Bing Feng and Dubey, Swapnil},
  journal={Applied Energy},
  volume={271},
  number={},
  pages={115147},
  year={2020},
  publisher={Elsevier} 
}

@article{maddalena2022experimental,
  title={Experimental data-driven model predictive control of a hospital {{HVAC}} system during regular use},
  author={Maddalena, Emilio T and Mueller, Silvio A and dos Santos, Rafael M and Salzmann, Christophe and Jones, Colin N},
  journal={Energy and Buildings},
  volume={271},
  number={},
  pages={112316},
  year={2022},
  publisher={Elsevier} 
}

@ARTICLE{7087366,
  author={Sturzenegger, David and Gyalistras, Dimitrios and Morari, Manfred and Smith, Roy S.},
  journal={IEEE Transactions on Control Systems Technology}, 
  title={Model Predictive Climate Control of a {{Swiss}} Office Building: Implementation, Results, and Cost–Benefit Analysis}, 
  year={2016},
  volume={24},
  number={1},
  pages={1-12}
}

@article{vsiroky2011experimental,
  title={Experimental analysis of model predictive control for an energy efficient building heating system},
  author={{\v{S}}irok{\`y}, Jan and Oldewurtel, Frauke and Cigler, Ji{\v{r}}{\'\i} and Pr{\'\i}vara, Samuel},
  journal={Applied Energy},
  volume={88},
  number={9},
  pages={3079--3087},
  year={2011},
  publisher={Elsevier} 
}

@article{vzavcekova2014towards,
  title={Towards the real-life implementation of {{MPC}} for an office building: Identification issues},
  author={{\v{Z}}{\'a}{\v{c}}ekov{\'a}, Eva and V{\'a}{\v{n}}a, Zden{\v{e}}k and Cigler, Ji{\v{r}}{\'\i}},
  journal={Applied Energy},
  volume={135},
  number={},
  pages={53--62},
  year={2014},
  publisher={Elsevier} 
}

\end{document}